\documentclass[aps, prb,reprint,10pt,superscriptaddress,amssymb,amsfonts]{revtex4-1}
\usepackage{amsmath,amssymb,amsthm,amscd,latexsym,epsfig,bm}

\usepackage[colorlinks,bookmarks=false,citecolor=blue,linkcolor=red,urlcolor=blue]{hyperref}
\usepackage{verbatim}

\begin{document}

\title{Local Master Equation for Small Temperatures}

\author{Evgeny Mozgunov}
\affiliation{Department of Physics and Institute for Quantum Information and Matter, California Institute of Technology, Pasadena, California 91125, USA}

\email[]{emozguno@caltech.edu}
%\homepage[]{Your web page}
%\thanks{}
%\altaffiliation{}
%\noaffiliation
\date{\today}

\begin{abstract}
We present a local Master equation for open system dynamics in two forms: Markovian and non-Markovian. Both have a wider range of validity than the Lindblad equation investigated by Davies. For low temperatures, they do not require coupling to be exponentially weak in the system size. If the state remains a low bond dimension Matrix Product State throughout the evolution, the local equation can be simulated in time polynomial in system size.
\end{abstract}

\maketitle

\section{Introduction}

%\textcolor{red}{[The introduction is collected from disconnected discussions]}

What are the effects in nonequilibrium dynamics where numerical simulation is more predictive than theory? A good example are nonlocal hops found during adiabatic evolution of a disordered system \cite{Vedika}. One can envision an experiment where this effect is measured in an ensemble of disordered systems. The results will be affected by interactions with the environment, so one needs to simulate open system dynamics to make any prediction of the results.
%This effect is observed in the adiabatic limit, so a good simulation tool should be able to access times long in comparison to characteristic timescale of the system.
Note that even the thermodynamics of disordered systems is hard for any analytic or numerical method, as it is a quantum version of classical glassy systems. These systems can encode NP-hard problems, and development of their theory is held back by intractability of a similar nature. A naive numerical study of their quantum behavior is bound to be exponentially hard in the system size. We believe that one can do better than that, at least for a special case of systems that are Many-body localized. In particular, we would wish to simulate quantum adiabatic annealing of such systems in the realistic setting that already has a lot of experimental data. \cite{Lidar,Tom}

%We seek to provide a window of numerical predictions into the world of quantum annealing of disordered systems.
Here we are developing numerical tools to realize this idea. We formulate a new Master equation that has local structure and is appropriate for simulating dynamics of states that have local structure as well (such as Many-body localized states\cite{Imbrie}). Previous work\cite{FV,CL,HMM} often used the same approach without stating it in a form of a new equation. Other equations have been proposed\cite{LidarWrites1,LidarWrites2,LidarWrites3}, but the locality has not been stressed out. The traditional Lindblad master equation at finite temperature loses all the locality, so it becomes incompatible with modern Matrix-Product State methods\cite{Monster, Vidal}. In this equation, a Rotating-wave approximation\cite{davies1974markovian} is made that leads to its loss of locality. Terms containing $e^{i(\omega-\omega')t}$ are dropped from the 
%right-hand side of the
Master equation with an assumption that $(\omega - \omega')\gg 1/T_1,T_2$.\footnote{This is stated most explicitly in Zoller, page 88, about eqn. 3.6.67}  Here $T_{1,2}$ are decoherence times, 
%and $T_{LS}$ is the time-scale corresponding to Lamb shift.
%Recall that $T_{1,2} \sim 1/Aw$. In formula below Aw =
So the coupling to the bath should be weaker than typical level spacing, which for $\|H\| =N$ qubits goes as $N/2^N$. For 10 qubits that would mean that $0.01 \gg 1/T_{1,2}, \quad T_{1,2} \gg 100$. Thus we need times of decay at least $1000$ in the units corresponding to the norm of local term in the Hamiltonian. This is grossly not true in systems like D-wave (time of decay is of order $1$). 

Local Master Equation appears at the intermediate stage of the derivation of traditional Lindblad form\cite{davies1974markovian}. We show that this equation possesses the same properties as the Lindblad one: positivity and Gibbs-preservation, but has wider range of applicability. 
%In the equation derived in this work, there's no need for Rotating-wave approximation, so one can study many-body systems (for which that approximation is not applicable). 
Note that any nonequilibrium study in the literature that was using the traditional Lindblad Master equation can be redone with the new Master equation, and different results can be found. The goal of this paper is to provide a proof of concept using the simplest possible examples.
%example of quantum adiabatic annealing.
Future applications include quantum adiabatic annealing, quantum thermodynamics\cite{Geva} and stability proofs for open system evolution\cite{Spy}.

%We want to simulate 1 qubit and multiqubit Quantum mechanics in contact with thermal environment.

%The traditional tool to do this is a Lindblad Master equation. The knowledge of the Hamiltonian eigenstates is needed for formulating of this Master equation. That doesn't make sense. How would the bath know the global eigenstates, if it only touches the system locally?

%Turns out this is due to a bunch of approximations done on the way. Most notably, Master equation is derived from coupling to oscillator bath, and a lot of 

%This means that the standard quantum optics derivation is not applicable. 

\paragraph{Local Master equation}

The evolution of the density matrix $\rho_s$ of a system with the Hamiltonian $H_s$ interacting with the environment via a single operator $A$ is given by:

\begin{align}
  \frac{d\rho_{s}}{dt} = - i[H_s,\rho_s(t)] + \frac{1}{3} (  [A^{f}\rho_{s}(t),A] +  [A,\rho_{s}(t)A^{{f}\dag}]+ \label{FinalForm}  \\
  +  [A^{f}(T')\rho_{s}(t),A(T')] +  [A(T'),\rho_{s}(t)A^{{f}\dag}(T')] \\
  +  [A^{f}(-T')\rho_{s}(t),A(-T')] +  [A(-T'),\rho_{s}(t)A^{{f}\dag}(-T')] )
\end{align}

 The filtered function is defined as:
\begin{equation}
  A^f = \int_{-x}^\infty d\tau C(\tau)A(-\tau)
\end{equation}
 where we can take $x$ either $-\infty$ and $0$. Taking $x=0$ will result in an evolution that does not preserve Gibbs state in general (which one partially fixes with counterterms), but it may preserve it in practice well enough.  $x=-\infty$ preserves the Gibbs state perfectly but has weaker error bounds due to other reasons.

 The bath correlation function is defined to be:
 \begin{equation}
   C(t) = N e^{-(t^2-i\beta t)/4t_{bath}^2} \label{corfunk}
 \end{equation}
 where $N$ is a normalization constant chosen such that $\|A\| \approx \|A^f\|$ for convenience. One can then rescale $A$ to vary the strength of the noise. $T'$ is the time-averaging window needed to restore positivity; in practice it is enough to set it to $0.3 \|H_{loc}\|^{-1}$ - norm of the local terms in the Hamiltonian.

\paragraph{Local Integral equation}
An alternative form that has the largest range of applicability is:

\begin{align}
  \frac{d\rho_{s}}{dt} = - i[H_s,\rho_s(t)] + \\
 [\int_0^\infty A(-\tau)C(\tau) \rho_s(t-\tau) d \tau ,A] + \\+ [A,\int_0^\infty A(-\tau)C^*(\tau) \rho_s(t-\tau) d \tau]-\\ - iD(0)( [A\rho_{s}(t),A] -  [A,\rho_{s}(t)A]) \label{FinalForm2} 
\end{align}
Its time-averaged form can be found in Appendix \ref{app:ta}. The $D(0)$ is an approximation to what's actually needed to complete the integral to $\int_{-\infty}^\infty$. Its value is:
\begin{equation}
    D(0) = -\frac{i}{2}\int_{-\infty}^\infty \textrm{sgn}(t) C(t) dt\text{.}
\end{equation}

In Section \ref{firstDeriv} we revisit the derivation of the above equation. In Section \ref{gbbs} we establish the Gibbs-preservation. In Section \ref{IE} we focus on integral equation. Then we proceed to study applications of the derived methods in Section \ref{apps}. The theoretical bounds on error are presented in Section \ref{errors}. We note the possibility of reduction from density matrices to states via the formalism of Quantum jumps in Section \ref{toinfinity}. Then in Section \ref{posit} we use the same formalism to convince ourselves in positivity of the evolution described by the new equation. 
\section{Derivation of Master Equation}
\label{firstDeriv}
The typical derivation of the Master equation starts from  
\begin{equation}
 d\rho/dt  =-i[H_s + V + H_b,\rho]   
\end{equation}
%sum down there
Take for simplicity $V =  A \otimes B$ where B is the bath operator, and both $A$ and $B$ are Hermitean. A general case $V=\sum_iA_i \otimes B_i$ can be treated in the same way. In the interaction picture the above equation has a formal solution 
\begin{equation}
 \rho_i(t) =\rho(0) -i\int_0^t [V_i(\tau),\rho_i(\tau)] d\tau   \label{Two}
\end{equation}
We can plug it into the equation: 
\begin{equation}
d\rho_i/dt  =-i[V_i(t),\rho(0)]-[V_i(t),\int_0^t [V_i(\tau),\rho_i(\tau)] d\tau] \label{integralOp}   
\end{equation}

So far all the steps were exact. So whatever integral operator stands in Eqn. (\ref{integralOp}), its application (e.g., by storing the result of the integration $\int_0^t [V_i(\tau),\rho_i(\tau)] d\tau$) gives completely positive evolution. We now assume that $\rho = \rho_s \otimes \rho_b$ at all times and take the trace. The positivity may be lost, but we operate under the assumption that it is still present. We will check that assumption when we arrive at the final form.
We arrive to 
\begin{align}
  \frac{d\rho_{i,s}}{dt} = \int_0^t d\tau C(t-\tau) [A_i(\tau)\rho_{i,s}(\tau),A_i(t)] +\\+ \int_0^t d\tau C^*(t-\tau) [A_i(t),\rho_{i,s}(\tau)A_i(\tau)]   
\end{align}
Where $C(t) = \textrm{tr}\rho_e B(t) B(0)$ is the bath correlation function to be discussed below. We did not assume anything about Markovianity.  The above integral equation is hard to solve because one needs to store $\rho(t)$ for all times. Indeed, the relative weight of time $\tau$ in the integral changes with $t$ according to $C(t-\tau)$. So there's no trick to simulate the equation above, one just needs to honestly store a lot of $\rho$'s.

The Markovianity assumption comes into play here. We assume that over times $t_{dec}$ when $C(t-\tau)$ is not too small, the change in $\rho_i$ induced by the coupling to the bath is small. So one neglects it to the first order and replaces $\rho(\tau) \to \rho(t)$. After that, the integrals can be moved as below:
\begin{align}
  \frac{d\rho_{i,s}}{dt} =  [\int_0^t d\tau C(t-\tau)A_i(\tau)\rho_{i,s}(t),A_i(t)] +\\+  [A_i(t),\rho_{i,s}(t)\int_0^t d\tau C^*(t-\tau)A_i(\tau)]   
\end{align}
or better still, we may introduced filtered operator (it will be nonhermitean):
\begin{equation}
  A_i^f(t) = \int_0^t d\tau C(t-\tau)A_i(\tau)
\end{equation}
Final assumption is to send one of the limits to $-\infty$
\begin{equation}
  A_i^f(t) = \int_{-\infty}^t d\tau C(t-\tau)A_i(\tau)
\end{equation}
 The equation assumes a form:
 \begin{equation}
  \frac{d\rho_{i,s}}{dt} =  [A_i^f(t)\rho_{i,s}(t),A_i(t)] +  [A_i(t),\rho_{i,s}(t)A_i^{f\dag}(t)]   \label{badLuckForm}
\end{equation}
Now by slight abuse of notation we define "Shroedinger picture" filtered operator:
\begin{equation}
  A^f = e^{iH_s t} \int_{-\infty}^t d\tau C(t-\tau)A_i(\tau)e^{-iH_s t}
\end{equation}
%(check signs).
Since $A_i(\tau) = e^{-iH_s t}Ae^{iH_s t}$, the dependence on $t$ drops after shifting the range of integration:
\begin{equation}
  A^f = \int_0^\infty d\tau C(\tau)A(-\tau)
\end{equation}

Now in the Master equation, we can go to Shroedinger picture as well:
\begin{equation}
  \frac{d\rho_{s}}{dt} = - i[H_s,\rho_s(t)] +  [A^f\rho_{s}(t),A] +  [A,\rho_{s}(t)A^{f\dag}]   
\end{equation}
Note that $A^f$ is approximately local for $C(t)$ decaying sufficiently fast due to Lieb-Robinson bound. Let's now consider a more general bath $V = \sum_i A_i \otimes B_i$. The equation takes the form:
\begin{equation}
  \frac{d\rho_{s}}{dt} = - i[H_s,\rho_s(t)] +  \sum_{ij}[A^f_{ij}\rho_{s}(t),A_j] +  [A_j,\rho_{s}(t)A^{f\dag}_{ij}]   
\end{equation}
where
\begin{equation}
  A^f_{ij} = \int_0^\infty d\tau C_{ij}(\tau)A_i(-\tau)
\end{equation}
and $C_{ij}$ is the bath correlation between $B_i$ and $B_j$. If those operators are local, one may use a bath that either have them completely independent, or such that correlations decay sufficiently fast. Then one can truncate $A^f_{ij}$ to $|i-j|<R$ without introducing big error into equation. The radius of locality of the generators in master equations thus increases by $R$ only.

We derived a local Master equation. There were only three assumptions along the way:
\begin{itemize}
    \item Separability, or Born approximation. One can check it by estimating the norm of the deviation from separability for any specific solution we find.
    \item Markovianity. If a solution is found, its characteristic timescale can be compared to the timescale of the bath $t_{dec}$
    \item Limits of integration. This one can mess up the first $t_{dec}$ of the solution, so, e.g. something that starts as $t^2$ will start as $t$, but after that time its effect is negligible.
\end{itemize}

There're two problems with the equation above. First, it needs to be time-averaged over a small window to restore positivity, as discussed in Section \ref{posit}. Second, for appropriate thermal correlation function, the Gibbs state is not a fixed point of the evolution. We will need to extend the limits of the integration in $A^f$ to
\begin{equation}
  A_{f0} = \int_{-\infty}^\infty d\tau C(\tau)A_i(-\tau)
\end{equation}
to make Gibbs state an exact fixed point, as shown in Section \ref{gbbs}. However, we see the deviation of the fixed point from Gibbs state as a feature of our model, not a bug. Indeed, original equation on the total density matrix
\begin{equation}
\rho_{tot}'(t) = i[\rho_{tot}, H_b +H_s +V]
\end{equation}
preserves the total Gibbs state $\sim e^{\beta(H_b +H_s +V)}$. It will preserve any diagonal in the total eigenbasis density matrix just as well. But it will {\emph{not}} preserve any product state $\rho_s \otimes \rho_b$ that is used in the Born approximation. So if we just look at this fact, we should not expect the Master equation to have a Gibbs fixed point. In Appendix \ref{true} we find that the true fixed point is, instead, the reduced density matrix of the total Gibbs state:
\begin{equation}
    \rho_{true ~ f.p.} = \frac{1}{Z} \textrm{tr}_b e^{\beta(H_b +H_s +V)}
\end{equation}
so our equation captures the deviations from thermodynamical predictions on the interface between the system and the environment.
\section{Gibbs state fixed point}
\label{gbbs}
  %Maybe it has fixed cycles or something like that. (?)

%Here's an extensive list of facts about linear systems and their stability:
%http://texas.math.ttu.edu/~gilliam/ttu/ode_pde_pdf/Ch4.pdf

%The main thing is to look at the eigenvalues. If in our $\rho' = M\rho$ equation some of the eigenvalues of $M$ (as $d^2$ by $d^2$ matrix) have positive real parts, then the system is unstable and should probably not be used. If some eigenvalues are purely imaginary, there are loops that do not fall to the fixed point. If more than one eigenvalue has zero real part, then there's not one fixed point. But if all of them have small negative real part (except for the one responsible for the trace, and one needs to be careful about complex conjugate matrix elements) - then there is a fixed point.

Often a desired property of the open system evolution is that the Gibbs state of an isolated system is a fixed point exactly. Even though it's not very physical to require that (as we discussed above), we can get the state $\sim e^{\beta H_s}$ as a fixed point of the end equation using a certain trick.

Let's see which terms of the equation favor which fixed point. The $i[\rho,H]$ term preserves everything diagonal in the eigenbasis of the isolated system. The form of filtered operator $A$
\begin{equation}
A_f = \int_{-\infty}^0 A(t) C(-t) dt 
\end{equation}
leads to the following terms
\begin{equation}
A_f = \sum_{mn} |m\rangle \langle n| A_{mn}(\pi S (E_{mn}) + i D(E_{mn})) = A_{f0} +A_D
\end{equation}
The $S$ is the bath spectral density defined as follows:
\begin{equation}
    C(t) = \int_{-\infty}^{\infty} S(\omega) e^{i\omega t}
\end{equation}
$S(\omega)$ obeys the thermal law $S(\omega) = e^{-\beta \omega}S(-\omega)$  as shown in Appendix \ref{app:b}. Note that $A_{f0}$ can be expressed via $A$ as follows:
\begin{equation}
  A_{f0} = \int_{-\infty}^\infty d\tau C(\tau)A_i(-\tau)
\end{equation}
So if we had these limits of integration,
%and in Master equation it leads to terms that preserve the Gibbs state. We include it in $A_{f0}$,
 we would only have $S$ in Master equation. The isolated Gibbs state is preserved by $A_{f0}$ terms:
\begin{align}
     \pi\sum_{k}A_{nk} A_{km} ( (S(E_{nk})\rho^{G,s}_k -S(E_{kn})\rho^{G,s}_n) -\\-(S(E_{km})\rho^{G,s}_m - S(E_{mk})\rho^{G,s}_k )) +\dots= \frac{d\rho^{G,s}_{nm}}{dt}
\end{align}
One may observe that they cancel each other as grouped.

Now $D$ is the remaining part.
\begin{equation}
     D(E_{nm}) = \textrm{p.v.} \int_{-\infty}^\infty \frac{S(\omega)d\omega}{\omega -E_{nm}}
\end{equation}
It does not possess the thermal law, therefore $A_D$ terms will not preserve the Gibbs state

In traditional Lindblad Master equation at finite temperature, the RWA is applied (rotating-wave approximation), which amounts to requiring $n=m$ in the sum, and $k=k'$ in terms like $A_{nk}^f \rho_{kk'}A_{k'n}$ for offdiagonal $\rho$. What remains of $A_D \rho A + A \rho A_D^\dag$ terms is 
\begin{equation}
 i\sum_{k}A_{nk} A_{kn} ( D(E_{nk})\rho_{kk}  -  D(E_{nk})\rho_{kk} ) =0
\end{equation}
$-AA_D \rho    -\rho A_D^\dag A$ give offdiagonal contributions:
\begin{equation}
 i\sum_{k}A_{nk} A_{kn} ( -D(E_{kn})\rho_{nm} +D(E_{kn})\rho_{mn})=i[\rho, \delta H_{diag}]_{nm}
\end{equation}
 So we can include them in the dynamics as the lamb shift corrections to energies in Hamiltonian. But then there's a mismatch: the Lindblad terms relax the system to Gibbs state of $H_s$, but the dynamics is rotating with $H_s + \delta H$. One can rely on "counter-terms" trick and include $-\delta H$ into $H$ from the start, and then neglect the difference in the Lindblad terms as it leads to $A_f(e^{\beta \delta H}-1) \cdot A$ kind of terms, and $\delta H$ itself is $\sim A^2$, so the correction is fourth order in $A$ - smaller than the precision we care about for our equation.

It is left to the reader to decide whether to use counter-terms in a given specific case. Formally, Lamb shift should be included, but sometimes we want to study the specific Hamiltonian, so we may decide that it {\emph{becomes}} such after contact with the bath has been established. %The discussion of counterterms for our equation can be found in the Appendix.

In our Local Master Equation, The Gibbs non-preserving corrections $[A_D\rho_{s}(t),A] +  [A,\rho_{s}(t)A_D^{\dag}]$ are not of the form $i[\rho, \delta H]$ anymore. But we note that we can approximate them by such, at least at low frequencies. Let's get more specific.

\begin{figure}[h!]
\centering
\includegraphics[scale=0.6]{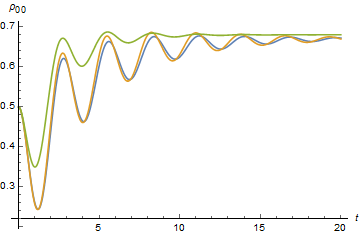}
\caption{ $\rho_{11}$ of single spin relaxation. Intergral equation (yellow)  vs. Markovian equation (blue) with the same order of Gibbs non-preservation  (with $\int_{-\infty}^0$ in $A_f$) and exactly Gibbs preserving (with $\int_{-\infty}^\infty$ in $A_f$) Markovian equation (green) }
\label{fig:cro1}
\end{figure}

$D(E)$ in our case will be the Dawson function, and it generally varies on the scale of the bath width.
Matrix elements of $A_D$  thus are not much different from matrix elements of $iD(0) A $ within that band. And higher matrix elements can be suppressed in systems that have exponentially small $A_{nm}$ for large energy change. So we make an approximation (an error of it is discussed in Section \ref{errors})
\begin{equation}
A_f \approx  A_{f0} +iD(0)A
\end{equation}
The second term then enters $\delta H = D(0) A^2$. Again, we can either simulate the dynamics of Lamb-shifted $H+\delta H$ system, or assume that the opposite shift was present in the isolated system. In the end, the equation with exact Gibbs preservation that we use for Fig. \ref{fig:cro1}-\ref{fig:eig} is the time-averaged version with all the counterterms gone:

\begin{align}
  \frac{d\rho_{s}}{dt} = - i[H_s,\rho_s(t)] + \frac{1}{3} (  [A^{f0}\rho_{s}(t),A] +  [A,\rho_{s}(t)A^{{f0}\dag}]+ \label{FinalForm111}  \\
  +  [A^{f0}(T')\rho_{s}(t),A(T')] +  [A(T'),\rho_{s}(t)A^{{f0}\dag}(T')] \\
  +  [A^{f0}(-T')\rho_{s}(t),A(-T')] +  [A(-T'),\rho_{s}(t)A^{{f0}\dag}(-T')] )
\end{align}

\begin{figure}[h!]
\centering
\includegraphics[scale=0.6]{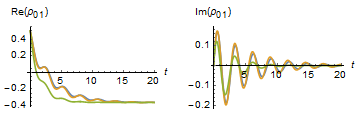}
\caption{$\rho_{12}$ of single spin relaxation. Intergral equation (yellow)  vs. Markovian equation (blue) with the same order of Gibbs non-preservation  (with $\int_{-\infty}^0$ in $A_f$) exactly Gibbs preserving (with $\int_{-\infty}^\infty$ in $A_f$) Markovian equation (green) }
\label{fig:cro}
\end{figure}

%\textcolor{red}{[We never really demonstrate Gibbs-preservation]}

\section{Integral Equation}
\label{IE}

%But for now, we note the following issue.
The Markovianity assumption required the bath time, which is at least $\beta$ (as shown in Appendix \ref{app:b}), to be smaller than the time when any significant change in $\rho$ happened. Even more strict than that, we would want to simulate a significant change in $\rho$, while constantly dropping contributions from its variation on the scale of $\beta$. If those variations are $\delta \rho \approx A^2 \beta$, neglecting them throughour the dynamics leads to error $ A^4 \beta T$. We know that $A^2T \approx 1$ since we want to see at least one half-life of the spin. So error $A^2 \beta \leq 0.01$. Which means $\beta \leq 0.01/A^2$. For the weak coupling, it seems to be an okay constraint but for a strong coupling we don't have hopes to access big $\beta$ (small temperatures). And that constraint is assuming that our counterterms worked out, in fact the precision of those operations may pose even stricter constraints on $\beta$. In particular, when we just use the "lookforward" approximation $\int_{-\infty}^\infty$ in $A_f$, it led to $A^2 \beta T \leq 0.01$ so $\beta \leq 0.01$.

In other words, for the problem at hand Markovianity is really out of question. One needs to see if there are non-Markovian integral equation that can still be simulated and possess the same nice properties: Gibbs-preservation, locality in some sense, and positivity.

The equation that we had before the Markovianity assumption was
\begin{equation}
    \rho' (t) = \int_0^t C(t-\tau)A(\tau)\rho(\tau) d\tau A(t) +\dots
\end{equation}
(the interacting picture is used here). The Shroedinger picture gives:

\begin{equation}
    \rho' (t) = i[\rho,H]+ \int_0^t C(\tau)A(-\tau)\rho(-\tau) d\tau A +\dots
\end{equation}

The upper limit of the integral can now be changed to $\infty$ as long we are sufficiently far into the evolution.
The closest we can get to Gibbs-preservation while maintaining accuracy is subtract $iD(0)A$:

\begin{equation}
    \rho' (t) = i[\rho,H]+ (\int_0^\infty C(\tau)A(-\tau)\rho(t-\tau) d\tau -iD(0)A \rho(t)) A +\dots
\end{equation}

The full time-averaged form is presented in  Appendix \ref{app:ta}. Generally speaking, there are better ways to restore the Gibbs fixed point. One notes that any equation of the form
\begin{align}
    \rho' (t) = i[\rho,H]+  \int_0^\infty C(\tau)A(-\tau)\rho(t-\tau) d\tau A +\\ +\int_{-\infty}^0 C(\tau)A(-\tau)\rho(t +f(\tau)) d\tau) A +\dots
\end{align}
will preserve the Gibbs state exactly. For $f(t)=0$ it is half-and-half combination of the Markovian and Non-Markovian equation. We do not know if such equations have any better bound on error if compared to original equation. Only in high-$T$ limit such bounds can be restored.

\section{Applications}
\label{apps}
\begin{figure}[h!]
\centering
\includegraphics[scale=0.4]{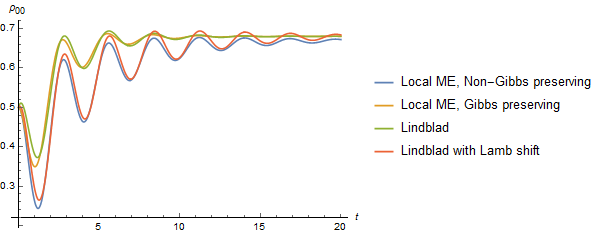}
\caption{Diagonal component $\rho_{11}$ of the density matrix 
%Real part (left) and Imaginary part (right).Some numerical instabilities kick in into the otherwise Hermitean matrix, but they remain small.
}
\label{fig:eig1}
\end{figure}

We start with a single qubit $|+\rangle \langle +|$ pure state, and simulate the Eq. (\ref{FinalForm111}) for time $T=100$ and $\beta =1$, which will be much bigger than what's allowed by our approximations, but so we can see how the method performs in practice. The Hamiltonian is now:
\begin{equation}
   H= 0.5 \sigma_z  +\sigma_x 
\end{equation}
and the $A=0.5\sigma_z$. We expect decay to $\beta =1$ fixed point
\begin{equation}
   \rho_{fp} = e^{-H}/\textrm{Tr}e^{-H}
\end{equation}
and indeed observe it to at least 6 digits of precision. The $\sqrt{\textrm{Tr}A^f A^{f\dag}}= 0.69$ now. The decay time is therefore expected to be $\|A\|\|A^f\|^{-1} \approx (0.5\cdot 0.69)^{-1} = 2.9$. We observed a value around that, with some oscillations. (See Fig. \ref{fig:eig1})

\begin{figure}[h!]
\centering
\includegraphics[scale=0.5]{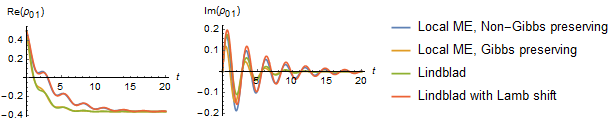}
\caption{Offdiagonal component $\rho_{12}$ of the density matrix, Real part (left) and Imaginary part (right).}
\label{fig:cro}
\end{figure}

\begin{figure}[h!]
\centering
\includegraphics[scale=0.4]{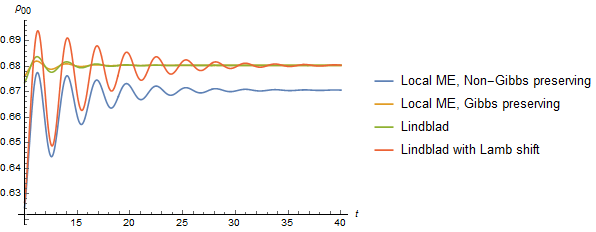}
\caption{Longer times, Diagonal component $\rho_{11}$ of the density matrix. We se relaxation to different fixed points. The fact that our equation captures the correct fixed point is explained in Appendix \ref{true}
%, Real part (left) and Imaginary part (right). Some numerical instabilities kick in in both equations into the otherwise Hermitean matrix, but they remain small.
}
\label{fig:eig}
\end{figure}

We now (In Fig. \ref{fig:eig1},\ref{fig:cro},\ref{fig:eig}) compare this evolution to an evolution with respect to traditional master equation:
 \begin{align}
  \frac{d\rho_{s}}{dt} = - i[H_s,\rho_s(t)] +\\+ \sum_{mn} [A^f_{mn}|m\rangle \langle n|\rho_{s}(t),A_{nm}|n\rangle \langle m|] +\\+  [A_{mn}|m\rangle \langle n|,\rho_{s}(t)(A^{f\dag})_{nm}|n\rangle \langle m|]   
\end{align}
where $|n\rangle,|m\rangle$ are eigenstates of the system Hamiltonian. The resulting evolution possesses the same features, however we find deviations of around $0.1$ in the oscillations before the equilibration.

%\begin{figure}[h!]
%\centering
%\includegraphics[scale=0.3]{OffdL.png}
%\caption{Discrepancy between two equations, traditional minus ours, Offdiagonal component $\rho_{12}$ of the density matrix, Real part (left) and Imaginary part (right).}
%\label{fig:cro}
%\end{figure}

\paragraph{Powder of sympathy} A good test for our ideas is the setup 
 \begin{equation}
     H = 100\sigma_1^z + \sigma_2^z + e^{-50} \sigma_1^x \sigma_2^x + 0.01 \sigma_2^x B
 \end{equation}
Where $B$ is a $T=1$ bath operator. The relaxation of the excited state of spin $1$ is all due to the bottleneck $e^{-50}$, so we don't expect to see anything up to exponential times. However, all our equations can produce faster relaxation if the bandwidth is broad enough, or $t_b<0.01$, and the relaxation of $\sigma_2^z$ has normal $T_1,T_2 \approx 100$. This provides more constraints on the bandwidth: it shouldn't be much bigger than $T$ or the energy scale of individual qubits, or it can lead to exponentially big bath fluctuations.

An interesting interpretation of this gedanken experiment is an old 17th century proposal of testing the effects of the "Powder of Sympathy", that can be found on the Wiki. %There's even a novel about it.

\paragraph{Locality} So far, we didn't gain much by switching to local equation - one needs to calculate every matrix element of $d$-by-$d$ matrix $A_f$, which can easily take longer than the evolution itself. But note that the expression 
 \begin{equation}
     A^f\rho_{s}A +  A\rho_{s}A^{f\dag} 
 \end{equation}
 only needs to be known up to some precision - accuracy of our evolution equation. So far we've made 3 (and more) approximations, now we're just going to make another one by neglecting exponential tails of $A^f$. We note that
 \begin{equation}
     A^f = A^f_c  + \epsilon, \quad \textrm{where} \quad A^f_c = \int_0^\infty d\tau C(\tau)e^{iH_c \tau}A_i e^{-iH_c \tau}
 \end{equation}
 and $H_c$ is the Hamiltonian truncated to a block $B$ around $A$: 
 \begin{equation}
     H_c = \textrm{tr}_{\overline{B}}H
 \end{equation}
 There are two contributions to $\epsilon$:
 \begin{equation}
     \|\epsilon_1\| \leq \|A\| \int_{T_x}^\infty C(t) dt
 \end{equation}
 for some $T_x$ such that $B/2 - \textrm{supp}A/2 -vT_x = x$. Here $v$ is the Lieb-Robinson velocity (essentially, norm of the Hamiltonian). The second contribution to error is then bounded by Lieb-Robinson bound:
 \begin{equation}
    \| \epsilon_2\| \leq \|A \|\int_0^{T_x} |C(t)| dt e^{-cx}
 \end{equation}
 where $c$ is a constant of order 1. What we derived is that we can approximate the timestep operators to a given error. In practice, we want $\epsilon<1/T$ where $T$ is the desired time we want to evolved the system for. For really short bath times the Lieb-Robinson bound may not even by tight. Anyways we can get away with ln$T$ blocks.
 
 In this way, we get the size of the block to be at least $v\beta$ (see decay time of the bath above). This illustrates the hardness of simulating low temperatures. Lower temperatures require us to compare energies at larger and larger scale.

\paragraph{Evolving MPS} 

Another question is whether implementing a difference scheme for local Master equation and Integral equation can be done locally. We assume that we only want to keep track of local operators. Then any state can be evolved for time $T$ with computation time exponential in $vT$ (where $v$ is the Lieb-Robinson velocity). That is achieved, e.g. just by truncation to a block of size $\sim vT$ around the operator whose dynamics we wish to simulate. Also, any system can be evolved for time $T$ with computation time polynomial in $T$ and exponential in the system size. If one wishes to have simulation polynomial both in system size and in evolution, one needs to rely on approximate an ansatz such as Matrix Product States (MPS). There are ways to truncate the entanglement generated during evolution, and do so efficiently by looking at the state data only locally. Some of these truncation schemes will introduce a large error in practice. The choice of best truncation scheme is a topic of current research. Let's assume that one chooses one of those schemes and is given a promise that it will succeed in approximating the evolution for a given system. After the efficient (poly-time) truncation, the evolution step is done. Its result can be represented as a sum of local operators acting on the MPS state. Then there are efficient ways of representing that sum as a new MPS. 

Let's illustrate that polynomial number of additions can be done on MPS states in poly-time. %Let's see if that actually works.
We act by a local operator on a state at $i_0$. This adjusts the tensor $A_{i_0}$. So we need to consider a sum of the original MPS and the adjusted one:
\begin{equation}
    \sum_{\sigma}\otimes_i A_{i\sigma} |\sigma\rangle + \sum_{\sigma}\otimes_{i<i_0} A_{i\sigma} |\sigma\rangle \otimes A_{i_0\sigma}' |\sigma\rangle \otimes_{i>i_0} A_{i\sigma} |\sigma\rangle =
\end{equation}
It's all linear, the matrix multiplications are implied. We can directly add them now:
\begin{equation}
   = \sum_{\sigma}\otimes_{i<i_0} A_{i\sigma} |\sigma\rangle \otimes (A _{i_0\sigma}+A_{i_0\sigma}') |\sigma\rangle \otimes_{i>i_0} A_{i\sigma} |\sigma\rangle 
\end{equation}
But then if we want to add a term with another $i_1$ adjusted, such nice form cannot be preserved anymore. However, if we assume that $A'$ is small, then preserving this form will only make an error of order $A'^2$ in any amplitude. We think that's good enough.

So far, we glossed over how fine the memory of $\rho(t)$ has to be stored for the calculation of integrals. Thanks to smoothing, we do not expect $\rho(t)$ to have any oscillations with frequencies higher than $1/T'$.
So that is a good interval to store values of $\rho$. But then one needs to be careful when actually integrating, as $A(t)$ as well as $C(t)$ may contain rapid components. Anyways, $C(t)$ can be truncated in time, and $A(t)$ can be truncated in space. This gives a poly-time evolution-step for MPS-like system, as long as the long-range entanglement is truncated after every step. Of course, we do not have a good reason to truncate the entanglement - one never knows when it will come back and interfere on itself. But as long as the small error is promised to us, the faithful low-T simulations can be done in poly-time.

%Let me put the basic (non-constructive) idea of truncation of entanglement here. Split the system into regions ABC, where B is a buffer of constant size. Apply independent unitaries on A and C so that one entangled pair is isolated across AC. Measure that pair, now invert the used unitaries. This is an elementary step of truncation.

%One can envision a procedure that checks whether such step is possible for a given state and division into regions. Then one scans the system and whenever step is possible, performs it, until all $S>1$ entanglement longer than B is removed.

%In practice, of course, there's no efficient way to check for above. However, assuming the finite speed of propagation, one can truncate regions A and C, and then we are dealing with finite-dimensional systems, so however inefficient the algorithm for scanning and the step may be, for the purposes of complexity it can still be applied in poly-time.

\section{Error bounds} 
\label{errors}
 Let's recap the timescales involved:
\begin{itemize}
    \item $t_{b}$ - the characteristic timescale of the correlator of the bath.
    \item $\beta$ - temperature can also be thought of in the units of time
%    \item $\Delta t$ - this time scale will not appear in the final equation, but was used for derivation. It's "lookforward time" that we allow for the Gibbs state to be a fixed point. We will need to arrange other timescales to make room for this one.
    \item $T'$ - the timescale used  for time-averaging
    \item $(\|A\|\|A^f\|)^{-1}$ - the decoherence time scale
    \item $\|H_{loc}\|^{-1}$ - inverse of typical internal frequency of our system
    \item $4^n \|H\|^{-1}$ - inverse of smallest level spacing
    \item $T$ - the desired evolution time, typically at least $(\|A\|\|A^f|)^{-1}$
\end{itemize}

Recall that $V = A\otimes B$ has dimension of energy. Thus 
\begin{equation}
    C(t) = \textrm{tr} \rho_{b,G} B(t) B
\end{equation}
has dimension of energy as well. Assuming:
\begin{equation}
    S(\omega) = e^{-\omega t_b^2 - \beta \omega /2}
\end{equation}
we get
\begin{equation}
    C(t) = \frac{\sqrt{\pi}}{t_b} e^{\frac{\beta^2/4 - t^2 +i \beta t}{4t_b^2}}
\end{equation}
that has the right dimension. It has its maximum absolute value at $t=0$:
\begin{equation}
  C(0) =\frac{\sqrt{\pi}}{t_b} e^{\frac{\beta^2}{16t_b^2}}  
\end{equation}
and width $\sim \beta \textrm{ln} \beta $ for $t_b = \beta/4$. We can make width even bigger for greater $t_b$, but it does not get much smaller for smaller $t_b$. Bigger width leads to weaker error bounds, so we fix $t_b =\beta/4$. As a warmup, we bound the norm of $A_f$:
\begin{equation}
  \|A_f\| = \|\int_0^\infty C(t) A(-t) dt \|\leq C(0) \beta  \|A\| = 4\sqrt{\pi}e\|A\|
\end{equation}
we neglected the logarithm. 
%We will drop the numerical factors as well in what follows.
The right-hand side of the equation is then of order:
\begin{equation}
  \|\dot\rho_i\| \leq 4\pi e \|A\|^2 \quad \Rightarrow \quad \|\Delta\rho_i\| \leq 4\pi e \|A\|^2t
\end{equation}
So a $O(1)$ change in $\rho_i$ will happen in time such that $\|A\|^2t=1$. We also note that the first $t\leq \beta$ of time are not very accurate for the Markovian equation, because integrals $\int_0^t$ were replaced by $\int_{-\infty}^t$. An integral equation does not suffer from that problem, if one keeps track of proper limits. So a meaningful comparison is possible only for $t\geq \beta$. We can actually make our Markovian equation work in that regime as well, by introducing time-dependent $A_f \sim \int_0^t$ in the beginning of evolution. 

But let's consider $t\geq \beta$ for simplicity. The rate of change in $\rho_i$ is $4\pi e \|A\|^2$, so the time-averaging replaces $\rho$ with something different by $\delta \rho \leq 4\pi e \|A\|^2 T_{ave}$ (In fact there are some cancellations, but considering them does not lead to a tighter bound). That difference leads to a  deviation in the evolution of order:
\begin{equation}
  \epsilon_{ave} = (4\pi e)^2 \|A\|^4 T_{ave} t
\end{equation}
In the Markovian equation, we also replaced $\rho(\tau)\to \rho(t)$ on scales of order $\beta$. In a similar way that leads to error
\begin{equation}
  \epsilon_{mkv} = (4\pi e)^2 \|A\|^4 \beta t
\end{equation}
Note that for $t\leq \beta$ it's more like $\|A\|^4 t^2$, if one is careful in defining time-dependent $A_f$. The Davies derivation does time-averaging as well, on a bigger scale $T_{ave}^*$. But it also drops some of the terms $A\to A-\delta A$ after it. The scale of those terms is \begin{equation}
 \| \delta A\| \leq \frac{\|A\|}{T_{ave}^* \delta E} d^a
\end{equation}
where the factor $d^a$ appears during the transition between bounds on individual matrix elements and bounds on the norm. $d$ is a dimension of a full system - we do not know how to do the truncation for the equation in Lindblad form. % or a block of size $\beta v \approx \beta \|H\|$ after the truncation, then $d$ is $2^{O(\beta \|H\|)}$.
The minimum level spacing is $\sim 1/d^2 $.
So even if there's a tighter bound for norm transition, the factor exponential in system size still remains. Two terms depending on $T_{ave}$ in the Davies equation are:
\begin{equation}
  \epsilon_{ave}  + \epsilon_{RWA} = (4\pi e)^2 \|A\|^4 T_{ave}^* t +4\pi e \frac{\|A\|^2t} {T_{ave}^* \delta E} d^a
\end{equation}
minimization of the error leads to
\begin{equation}
  \epsilon_{ave}  + \epsilon_{RWA} = (4\pi e)^{3/2} \|A\|^3  t \sqrt{\frac{d^a}{ \delta E} }
\end{equation}
which is exponential in the system size.

Finally, we need to estimate the error of the Born approximation. Unlike the above estimates, we don't really have a good control over it. So instead of an explicitly rigorous bound (like that in Davies paper), we just provide a proxy for it. We note that during the derivation, we introduced a nesting in the equation, by plugging in a formal solution:
\begin{equation}
  \dot\rho_i(t) = i[V(t),\rho_i(0)] -[V(t),\int_0^\infty[V(\tau),\rho_i(\tau)]d \tau]
\end{equation}
We note that this formal solution that we plugged in does not exactly satisfy the Born approximation, even if $\rho_i(\tau)$ that's inside of it is Born. So one can imagine that by repeated nesting one improves the non-Born properties of the solution. Consider nesting two more times, and then taking the trace over bath. We get some equation on $\rho_{i,s}$, that schematically can be written as follows:
\begin{equation}
  \dot\rho_{i,s} = \int \int \int A^4 C^2 \rho_{i,s}
\end{equation}
Alternatively, we can get the equation of the same order by nesting the traced form that we used:
\begin{equation}
   \dot\rho_{i,s} = [\int A C \rho d\tau, A] +\dots \quad \Rightarrow \dot\rho_{i,s} = \int \int \int A^4 C^2 \rho_{i,s}
\end{equation}
The two results will not coincide, even though they have the same order of operators. So the first order of error in Born approximation %(which corresponds to crossing diagrams)
is
\begin{equation}
   \|\delta \rho\| \leq \|\int \int \int A^4 C^2 \rho_{i,s}\|t \sim \beta^3 \|A\|^4 \frac{1}{\beta^2}t=\|A\|^4 \beta t
\end{equation}
for $t\geq \beta$, and $\|A\|^4 t^4/\beta^2$ for $t\leq \beta$. It is essentially the same bound for $t>\beta$. Maybe the special structure of the correction to Born approximation leads to cancellations that improve the above bound, but we do not investigate this further. We present two pictures of ranges of validity, one with the Born approximation proxy, and another without it (just other errors).

Conclusion: for 1 qubit with $\|H\|=1$, the theoretical error bounds are summarized in the table:

%the local Markovian equation is valid for $\beta \leq 0.1/A^2$ {\color{red} wow wow wow} and the integral equation is valid for all $\beta$, as long as $A$ is sufficiently small for Born approximation to remain valid. The $t_b =\beta /4$.

%If we compare this to the original Lindblad, that one had $1/(A^2\textrm{max}(t_b, \beta)) =\Delta t \geq 1/\textrm{min}E_{nm} \sim 2^n$ for the full RWA time-averaging, where $n$ is the system size. So  $\beta \leq 2^{-n}/A^2$. This means exponentially large temperatures or exponentially small coupling. By truncation $n\to v \beta  +$ const  we see that it's not actually that bad of a constraint.
%\begin{equation}
%    \beta e^{v\beta} \leq \textrm{const}/A^2
%\end{equation}
% so $\beta =1/v \approx 1$ for $\|H\| =1$ is doable, but beyond that the $A$ has to become exponentially weaker. Let's say that const $=0.01$.

% All the previous constraints apply.
% We can summarize our results in the following table for $\|H\|=1$:(single qubit):

\begin{center}
\begin{tabular}{ |c|c|c| } 
 \hline $\epsilon/A^2t$ &  Born  & Other errors  \\ \hline
  Lindblad &  $A^2\beta$ & $A^2\beta + A$ \\ 
 Local ME & $A^2\beta$ & $A^2(\beta + 0.3)$   \\ 
 Integral equation & $A^2\beta$ & $0.3A^2$ \\ 
 \hline
\end{tabular}
\end{center}
These expressions are exactly the error rate per half-life $A^2t=1$. The $t>\beta$ is assumed. Most of the numerical factors are dropped for brevity, as bounds are not very tight in practice anyways.

\begin{figure}[h!]
\centering
\includegraphics[scale=0.6]{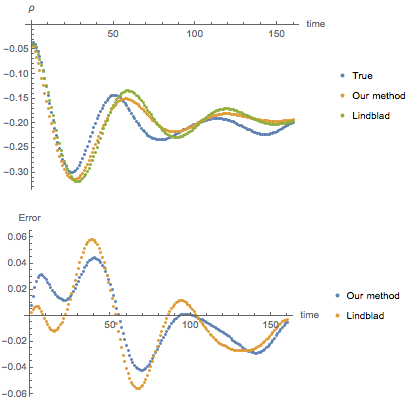}
\caption{Comparison with the faithful simulation of 8-qubit bath}
\label{crush}
\end{figure}

Now for $n$ qubit system with $\|H_{loc}\|=1$, the biggest difference is that now Lindblad error bound grows exponentially with the dimension. We also assume that there are multiple $A_i$ coupled to independent baths acting on each qubit, which leads to some polynomial factors as well:
\begin{center}
\begin{tabular}{ |c|c|c| } 
 \hline $\epsilon/A^2t$ &  Born  & Other errors  \\ \hline
  Lindblad &  $n^2A^2\beta$ & $n^2A^2\beta + n\sqrt{n}e^{cn}A$ \\ 
 Local ME & $n^2A^2\beta$ & $n^2A^2(\beta + 0.3)$   \\ 
 Integral equation & $n^2A^2\beta$ & $0.3n^2A^2$ \\ 
 \hline
\end{tabular}
\end{center}
There is a truncation scheme outlined in Section \ref{apps}. If one uses local forms of $A$, the for $t \lesssim \beta$ the dynamics is well approximated locally. $n$ can be replaced by $v\beta\approx \beta$ for $\|H_{loc}\|=1$. We get the best error bounds for $t_*=\beta$ :
\begin{center}
\begin{tabular}{ |c|c|c| } 
 \hline $\epsilon_{loc}/A^2t_*$ &  Born  & Other errors  \\ \hline 
 Local ME & $A^2\beta^3$ & $A^2(\beta^3 + 0.3\beta^2)$   \\ 
 Integral equation & $A^2\beta^3$ & $0.3A^2\beta^2$ \\  \hline
\end{tabular}
\end{center}
This is the error of local density matrices. We conclude that both of our equations avoid exponentially big $e^{cn}$ factor, and locally even the polynomial factors. One should keep in mind that these theoretical bounds are not very tight, and only kick in for big systems though. For just one qubit, the equations track each other almost perfectly well outside the regime of validity prescribed by the bounds above. Fig. \ref{crush} shows the result of faithful simulation of the bath of 8 qubits and 1-qubit system. Both methods lose to finite-size effects very fast.

Now we will see what are the extra requirements on $\beta$ that come from demanding Gibbs-preservation requirement as in Section \ref{gbbs}. %and it may well be possible that it will allow us to simulate low temperatures for some choice of bath.

%Investigation of Dawson function leads to following results:
The form of non-Gibbs preserving $D(E)$ function for $S(\omega) = e^{-(\omega t_b)^2-\beta \omega/2}$ 
%($\pi$ forgotten) 
is:
\begin{align}
    D(E) = e^{(\beta/4t_b)^2} \textrm{p.v.} \int_{-\infty}^\infty dx \frac{e^{-x^2}}{x- (Et_b -\beta /4t_B)} =\\= - e^{(\beta/4t_b)^2} \textrm{Dawson} (Et_b - \beta /4t_b)
\end{align}
It is a function with max height $\sim e^{(\beta/4t_b)^2}$ and characteristic width $1/t_b$ in $E$. So the derivative is bounded by $t_b e^{(\beta/4t_b)^2}$. A numerical investigation shows that one can plant $E=0$ on one of the extrema of Dawson function for $\beta/4t_b \approx 1$. Then the second derivative $t_b^2 e^{(\beta/4t_b)^2}$ is what gives the scale of Gibbs non-preserving terms at energy $E$:
\begin{equation}
    \rho_{mk}' \sim \sum_{mnkk'} A_{mn}A_{n'k} \rho_{nn'} (D(E_{mn}) - D(E_0))
\end{equation}
We note that for $E_{mn}<0$ the analogous terms with $S(E_{mn})$ are exponentially big, so we only need to worry about $E_{mn}\gtrsim-\beta^{-1}$ or so. 

%that we don't have to use $E_0 =0$ for our counterterms. The average rate of change in $\rho_{mk}$ is then:
%\begin{equation}
%    |\rho_{mk}'| \lesssim N_0|A_{m0}A_{0k}  (D(E_{m0}) - D(E_0))|
%\end{equation}
%where $0$ stands for a typical state in the lowest $T$ band of energies, and $N_0$ is the number of states there.
%We will be satisfied with our accuracy if the above rate is slower than other things that happen to $\rho_{mk}$ in our equation. For Markovian one, terms like $N_0 A_{m0}A_{0k} S(E_{0m})$ are actually very big as soon as the state $m$ is out of thermal band, because $S(E_{0m})$ contains exponential factor. So we only really need to care about $m$ within the thermal band, so $|E_{0m}|\sim T$. {\color{red}IDK about that} In that range, the error can be bounded:
The difference is then bounded
\begin{equation}
  |D(E_{mn})-E(E_0)|\lesssim  \frac{1}{16}(\frac{4t_b}{\beta})^2 e^{(\beta/4t_b)^2} \beta^2|E_{mn}-E_0|^2
\end{equation}
If we only want to suppress $\beta^2|E_{mn}-E_0|^2\lesssim 1$ terms, which would allow us to approximate well for $E_{mn}\lesssim \beta^{-1}$, we can maintain control over $D$.

%\begin{align}
 %   |\rho_{mk}'| \leq \approx |A_{m0}A_{0k}| t_b^2 e^{(\beta/4t_b)^2} |E_{0m}|^2 =\\= |A_{m0}A_{0k}| \frac{1}{16}(\frac{4t_b}{\beta})^2 e^{(\beta/4t_b)^2} |E_{0m}|^2
%\end{align}
The expression on the right for $\beta \Delta E =1$ has minimum at $t_b = \beta/4$:
\begin{equation}
    |D(E_{mn})-E(E_0)| \lesssim   \frac{e}{16}\beta^2|E_{mn}-E_0|^2 
\end{equation}
%\begin{equation}
%    |\rho_{mk}'| \leq \approx  |A_{m0}A_{0k}| \frac{e}{16} 
%\end{equation}
%we can say that $e/16 \ll 1$ so we prove that our evolution is approximately Gibbs-preserving modulo the counterterms. In fact, the exact matching to the extremum of Dawson happens very near to $t_b = \beta/4$. The form is independent of $\beta$, so the Gibbs-preservation actually does not lead to any further constraints on $\beta$!

\begin{figure}[h!]
\centering
\includegraphics[scale=0.6]{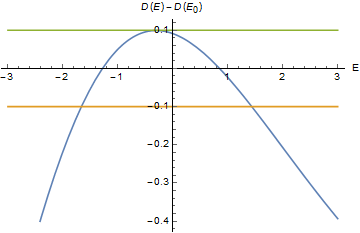}
\caption{Scale of the non-Gibbs preserving terms as compared to the terms we leave in the equation (%just $|A_{m0}A_{0k}|/A^2$
for $E_{mk}=E$ and $\beta=1$)}
\label{just}
\end{figure}

The Fig. \ref{just} shows a plot of magnitude of Gibbs non-preserving terms as a function of $E$ in units of $|A_{mn}A_{n'k}|$. We also use time averaging with $T'=0.1$, but since it corresponds to high frequencies $\geq 10$ it doesn't really matter.

We see that the magnitude is within 10$\%$ lines for $E \in [-1.5,1.5]$, for $T= 1/\beta = 1$. We also note that the picture is unchanged by rescaling $\beta \to n \beta$, $t_b \to n t_b$, $E_0 \to E_0/n$ and  $E \to E/n$. If for some reason we decide to ignore the deviations outside the $\sim T$ band, or $A$'s are such that suppress higher transitions, this is sufficient for Gibbs-preserving equation to be related to the true evolution.

\section{To Stochastic Shroedinger eqn}
\label{toinfinity}
Now we consider going from Master equation to Stochastic Shroedinger equation.
Master equation is a linear differential equation on $\rho$. We can write it concisely as
\begin{equation}
    \frac{d \rho_{ij}}{dt} =\mathcal{E}_{ijkl} \rho_{kl}
\end{equation}
A formal solution is then
\begin{equation}
    \rho_{ij}(t) = M_{ijkl}(t) \rho_{kl} = (e^{\mathcal{E}t})_{ijkl} \rho_{kl}
\end{equation}
(for time-dependent Hamiltonian there are extra complications) These equations describe deterministic evolution of a $d^2$-dimensional "vector" $\rho$. We can replace it by stochastic evolution of a $d$-dimensional object $|\psi\rangle$. In some cases only a value of an observable is requested at the end of the evolution, so it may be faster to sample from stochastic evolution of $d$-dimensional object than to run the full evolution of a $d^2$-dimensional object. It is also more straightforward to implement the wavefunction evolution of the MPS than to implement the MPO evolution of the density matrix (there are plenty of problems with preserving Hermiticity etc. of $\rho$ locally in the latter).

Formally, the equivalence goes as follows: consider a matrix $\rho'$ defined as $\rho'_{ik, jl} =\frac{1}{d} M_{ijkl}$. In case $M$ is a completely positive trace preserving map, $\rho'$ is a proper density matrix on two copies of the system ($d^2$-dimensional Hilbert space), as can be confirmed by acting with $M$ on one part of maximally mixed state in that Hilbert space. This gives us a way to check for positivity, if we're afraid we've lost it in our approximations. $\rho'$ is Hermitean (as long as equation on $\rho$ is invariant under Hermitean conjugation), so it can be diagonalized:
\begin{equation}
    \rho' = U D U^\dag
\end{equation}
Where $D$ is diagonal matrix. Moreover, if positivity of the evolution holds, $D\ge 0$, so one can write
\begin{equation}
    \rho' = \sqrt{\rho'} \sqrt{\rho'} = U \sqrt{D} \sqrt{D} U^\dag
\end{equation}
In indices, denote $\sqrt{\rho'}_{ik,mn} = N_{ik}(mn)$ (Alternatively, we can use $U \sqrt{D}$ for $N$, the result will be the same. In fact, there's freedom in the choice of $N$). Note that $N_{mn}(ik)= \sqrt{\rho'}_{mn,ik} = \sqrt{\rho'}_{mn,ik}^* =N_{ik}(mn)^*$ We get
\begin{align}
    \rho_{ij}(t) = M_{ijkl}\rho_{kl}  =\\= d \cdot \rho'_{ik,jl} \rho_{kl}= d \cdot N_{ik}(mn) N_{mn}(jl) \rho_{kl}  =\\= \sum_{mn}\sqrt{d} N_{ik}(mn) \rho_{kl} \sqrt{d} N_{jl}^*(mn)
\end{align}
We've made the summation over $mn$ indices explicit to stress out the point that our channel is a probabilistic sum:
\begin{equation}
    \rho(t) = \sum_{mn}\sqrt{d} N(mn) \rho \sqrt{d} N^\dag(mn)
\end{equation}
The normalization of these operators may be all over the place. $\sqrt{d}N(mn)$ are just complex $dxd$ matrices. They can map a normalized state $|\psi\rangle$ to something with norm less than $1$ or greater than one. The only constraint is that the total density matrix preserves its trace, so the norms should sum to $1$ over indices $mn$. 

To use this form for computation, one first needs to represent $\rho = \sum_i p_i |\psi_i\rangle \langle \psi_i|$,  then sample from $\{p_i\}$ and choose a corresponding $|\psi_i\rangle$. Then pick an integer $s = \{mn\}$ uniformly in $1\dots d^2$ and calculate $|\psi_i(t,s)\rangle = d\sqrt{d}N(mn)|\psi_i\rangle$. This unnormalized state is then used to calculate, e.g., expectation values of operators of interest.
\begin{equation}
    A(t,i,s) = \langle \psi_i(t,s)| A | \psi_i(t,s)\rangle
\end{equation}
These expectation values are averaged over random picks above.
\begin{equation}
    A(t) =\sum_i p_i\frac{1}{d^2}\sum_s \langle \psi_i(t,s)| A | \psi_i(t,s)\rangle
\end{equation}
One can easily check that
\begin{equation}
    A(t) = \textrm{tr}\rho(t)A
\end{equation}
as it should be. Quite a lot of distributions $P(A(t) =A(t,i,s)) = p_i/d^2$ are smooth enough so sparse sampling is sufficient.

Now we exploit the structure of our equation to come up with simple enough local operators $N$. The time step $dt$ is given by
\begin{align}
  \rho_{s}(t+dt) = e^{- iH_sdt}\rho_s(t)e^{ iH_sdt} +O(dt^2)+\\ +  (A^f\rho_{s}(t)A +  A\rho_{s}(t)A^{f\dag} -AA^f\rho_{s}(t) -  \rho_{s}(t)A^{f\dag}A )dt
\end{align}
The last two terms can be included as non-unitary part of the evolution:
\begin{align}
  \rho_{s}(t+dt) = e^{- iH_sdt -AA^fdt}\rho_s(t)e^{ iH_sdt -A^{f\dag}Adt} +\\+O(dt^2) +  (A^f\rho_{s}(t)A +  A\rho_{s}(t)A^{f\dag} )dt
\end{align}
Or, denoting the new, nonunitary evolution by $V$,
\begin{equation}
    \rho_{s}(t+dt) = V\rho_s(t)V^\dag +O(dt^2) +  (A^f\rho_{s}(t)A +  A\rho_{s}(t)A^{f\dag} )dt
\end{equation}

The right hand side needs to be represented as a sum of positive operators:
\begin{equation}
  V\rho_s(t)V^\dag + dtA^f\rho_{s}A +  dtA\rho_{s}A^{f\dag} = \sum_i C_i \rho_s C_i^\dag
\end{equation}

Note that the two terms in front of $dt$ for completely arbitrary hermitean $A$ and nonhermitean $B$ 
\begin{equation}
  B\rho_{s}A +  A\rho_{s}B^{\dag}
\end{equation}
is not necessarily a positive operator. In other words, this operation on $\rho_s$ is not necessarily a positive map. Indeed, if $A=1$ and $B=\sigma^+$, then the operation acting on a valid density matrix $|0\rangle \langle 0|$ results in a nonpositive operator $\sigma_x$.

So to find $C_i$,
 consider an operator-state mapping defined as follows:
\begin{equation}
    |A\rangle  =\sum_{ij}A_{ij}|ij\rangle, \quad |V\rangle = \sum_{ij}V_{ij}|ij\rangle \quad  \quad \langle V| = \sum_{ij}(V^\dag)_{ji}\langle ij |
\end{equation}

We arrive at the following form of $\rho'$:
\begin{equation}
\rho' = |V\rangle \langle V| + dt|A\rangle \langle A^f| + dt  |A^f\rangle \langle A|   
\end{equation}
We need to diagonalize it. Note that $\rho'$ acts as zero outside the span of vectors $|V\rangle, |A\rangle, |A^f\rangle$ in $d^2$-dimensional space. So all we need to deal with is 3x3 matrix (if there are $n$ operators $A$, then $(1+2n)$x$(1+2n)$). 
We can try to express the entries of the 3x3 matrix explicitly via the form of $\rho'$ and the Gram matrix of inner products of $|V\rangle, |A\rangle, |A^f\rangle$. For that we will need to choose a basis, e.g. one vector along $V$, second one perpendicular to it so $A$ lies in plane of these two, etc.

But we don't need to work with the orthogonal basis. Instead, consider the eigenvalue equation:
\begin{equation}
    \rho' |\lambda\rangle = \lambda |\lambda\rangle
\end{equation}
Call the vectors $|V\rangle, |A\rangle, |A^f\rangle$  by $|V_i\rangle$ where $i=1,2,3$. We assume they are linearly independent, and anyways
\begin{equation}
    |\lambda\rangle = \sum_i c_i|V_i\rangle
\end{equation}
with unique choice of $c_i$. Introduce the Gram matrix:
\begin{equation}
    G_{ij} =\langle V_i | V_j\rangle = \textrm{tr}V_i V_j^\dag 
\end{equation}
We arrive to the equation on vector $c$:
\begin{equation}
    \rho'Gc = \lambda c
\end{equation}
here $\rho'= (100,00dt,0dt0)$. So we have proven that non-Hermitean matrix $\rho'G$ has maximal number of independent eigenvectors. Moreover, the corresponding eigenvalues are real. We can just use the numerical procedure to deal with non-Hermitean matrix and it is guaranteed to find the full set.

Now using the values of $c$'s found numerically, we get
\begin{equation}
    \rho' = \sum_\lambda \lambda \sum_i c_i^\lambda |V_i\rangle \sum_j c_j^{\lambda*} \langle V_i| \lambda
\end{equation}
 For the moment here we assume $\lambda$'s will be positive which signals about the positivity of the map we're trying to simulate. In practice we will need to extend to bigger matrices and drop some terms to achieve it, which is described in the next Section.
 
 We translate back to jump operators by:
 \begin{equation}
     C_\lambda = \sqrt{\lambda}\sum_i c_i^\lambda V_i
 \end{equation}
 and
 \begin{equation}
     \rho_{s}(t+dt) = \sum_\lambda C_\lambda \rho_s(t) C_\lambda^\dag +O(dt^2)
 \end{equation}
 
 So the state should choose one of three paths with probabilities given by $\lambda$'s and transform as:
 \begin{equation}
     |\psi\rangle \to (c_1^\lambda V + c_2^\lambda A +c_3^\lambda A^f)\psi
 \end{equation}
 Note that $e^{iHt}$ can be implemented by a finite-depth circuit. $A$ is usually local, so we only need to figure out how to achieve locality of $A^f$, and then such jump can be calculated in $poly$-time for an MPS-like scheme.
 
 Aside: if for many $A$'s we want to have jumps associated with each $A$ to be separate, we may use splitting of the identity trick to repeat the above for each $A$ with pieces of $V$. 
 %described in the previous section.
 That will add a requirement $dt<O(1/n)$ on the time step. However, having them all together does not lead to any problems.
\section{Positivity}
\label{posit}
 The equation as it was derived originally is as follows:
 \begin{equation}
  \frac{d\rho_{s}}{dt} = - i[H_s,\rho_s(t)] +  [A^f\rho_{s}(t),A] +  [A,\rho_{s}(t)A^{f\dag}]  \label{dang}  
\end{equation}
The Eq. (\ref{dang}) 
is not guaranteed to give positive evolution. One may note, hovewer, that for $H=0$ $A_f\sim A$, and the positivity is restored. Another way to restore it is dropping terms like:
\begin{equation}
    \sum_{mn\ne lk}A_{mn}^f \rho A_{kl}
\end{equation}
which gives the standard ME in the Lindblad form, so explicitly positive. %(we will learn how to check positivity during the derivation of the Stochastic Shroedinger equation below).

One may ask what physical meaning is there in dropping the terms $mn \ne lk$? It can be achieved by smoothing the evolution. In equation 
\begin{equation}
     \frac{d\rho_{i,s}}{dt} = \int_0^t d\tau C(t-\tau) [A_i(\tau)\rho_{i,s}(\tau),A_i(t)] + \dots
\end{equation}
We can replace r.h.s.$(t)$ by $ \frac{1}{2T'}\int_{t-T'}^{t+T'}dt'$r.h.s.$(t')$. And then we still put $\rho(\tau) \to \rho(t)$. This averaging will make the terms $A_{mn}A_{kl}^f \sim O(1/T'|E_{mn}-E_{kl}|)$ if $|E_{mn}-E_{kl}|> 1/T'$. For a random  Hamiltonian of norm $\|H\|$ on $n$ qubits, the smallest $|E_{mn}-E_{kl}|\sim \|H\|/4^n$, so we need $T'\gg 4^n/\|H\|$ to arrive at the usual Lindblad form. Of course it is not desirable to average our equation over such long timescale - no dynamics will be left. There are certain tricks one can do with Lieb-Robinson bound and truncation that may allow one to do better than this. We also note that Lindblad form is guaranteed to be positive, whereas our equation does not. Positivity is restored for some $T'$, so this is a way to tweak our equation if it's not positive from the start in a given special case.

We find out that such evolution has third eigenvalue of the corresponding $\rho'$ (via channel-state duality) always negative. We fight that by time averaging the interaction part. We take a simple 3 point average:
 \begin{align}
  \frac{d\rho_{s}}{dt} = - i[H_s,\rho_s(t)] + \frac{1}{3} (  [A^f\rho_{s}(t),A] +  [A,\rho_{s}(t)A^{f\dag}]+ \label{FinalForm}  \\
  +  [A^f(T')\rho_{s}(t),A(T')] +  [A(T'),\rho_{s}(t)A^{f\dag}(T')] \\
  +  [A^f(-T')\rho_{s}(t),A(-T')] +  [A(-T'),\rho_{s}(t)A^{f\dag}(-T')] )
\end{align}
here $T'$ is a new time scale, in practice it is enough to choose $T' \approx 0.3 \|H_{loc}\|^{-1}$. For $\beta =1$ and $H= 0.5 \sigma_z + \sigma_x$ and $A= 0.5\sigma_z$ (something generic, but relevant for the D-wave problem) one obtains $T'=0.3$ as the averaging time when the negative eigenvalue becomes the 4th. Same $T'$ works for 2,3,5 qubit system with random chain Hamiltonians.

We note that such $T'$ is a big achievement of our method, the original Lindblad-Davies derivation needed $T'> 4^n/\|H\|$ the inverse smallest level spacing. % we compare the results of our evolution with the traditional Master equation. We also compare different versions of our method.

The way the positivity is checked is by investigating a $d^2$-by-$d^2$ operator, which actually has only $7$-by-$7$ matrix of nontrivial components (one for identity and two for each point in our averaging):
\begin{align}
\rho' = |V\rangle \langle V| +\frac{dt}{3} (|A\rangle \langle A^f| + |A^f\rangle \langle A| +  \label{qForm}  \\ + |A(T')\rangle \langle A^f(T')| + |A^f(-T')\rangle \langle A(-T')| +\\ + |A(-T')\rangle \langle A^f(-T')| + |A^f(-T')\rangle \langle A(-T')| )
\end{align}
here $|V\rangle = \sum_{ij}V_{ij}|ij\rangle$ on the two copies of a system, and $V = e^{-iHdt - AA^fdt}$. The bra-ket notation as applied to operators is somewhat misleading. The difference between the operator space and the quantum state space is that there's no good inner product on operator space. The above bra-ket imply the Tr$AB^\dag$ inner product, but it doesn't work out in quite the same way as the inner product for states.

A more mathematically correct way of posing the question is that $\rho'$ is a quadratic form $Q(O_1,O_2^\dag)$. It takes as input operators in span of $V,A,A^f\dots$ as one argument, their conjugates as another, and spits out a (generally complex) number. The quadratic form is completely defined by the Eq. (\ref{qForm}). For example, $Q(A,A^{f\dag}) = \frac{dt}{3}$. The value of $Q$ on any two operators $O_1,O_2$ can be expressed via linearity. Anything outside the span of our seven vectors is set to zero. If some of the vectors are linearly dependent, one chooses a basis. The resulting form output on two vectors is invariant with respect to a choice of basis.

Then one can ask to make this quadratic form diagonal. Once it's done, one investigates the values on the diagonal and checks for their positivity.

What we do is finding eigenvalues of $\rho'$ instead. Note that if $|\lambda\rangle  =\sum_{i}c_i |V_i\rangle$, then we arrive to equation on $c$ and $\lambda$:
\begin{equation}
    \left( \begin{array}{ccccccc}
1 & 0 & 0 & 0 & 0 & 0 & 0\\
0 & 0 & dt/3 & 0 & 0 & 0 & 0\\
0 & dt/3 & 0& 0 & 0& 0 & 0\\
0  & 0 & 0& 0 & dt/3 & 0 & 0\\
0 & 0 & 0& dt/3 & 0& 0 & 0\\
0  & 0 & 0 & 0 & 0& 0 & dt/3\\
0 & 0 & 0& 0 & 0& dt/3 & 0
\end{array} \right)Gc = \lambda c \label{BigMat}
\end{equation}
where $G$ is the matrix of inner products in the Tr$AB^\dag$ sense. One can rigorously derive the above equation. Consider acting by the evolution map on the maximally mixed state of two copies of the system (that is a pure state as a whole). The resulting density matrix is hopefully positive and surely Hermitean, so it has its eigenvalues. Its eigenvalue equation leads to Eq. (\ref{BigMat}).

So we are choosing the averaging timestep $T'$ so that negativity of matrix $\rho'$ for our actual map is not too big. Where does the negativity come from? Each 2-by-2 matrix $(dt/3)|A\rangle \langle A^f| + |A^f\rangle \langle A|$ is bound to have a negative eigenvalue. The eigenvalues are $\pm dt/3$(assuming a good normalization) if $\langle A^f|A\rangle = 0$ and $0,dt/3$ if $|A\rangle = |A^f\rangle$ . Finite temperatures are all somewhere in between these two cases. Now adding the term $|V\rangle \langle V|$ may in principle lift these two, but in practice overlap between $V \approx id$ and $A = \sigma_z$ is too small to do anything. So we do the averaging, going to 7-by-7 matrix. Now there are 7 eigenvalues, 6 of which form a block relating to $A$. This block for small $T'$ will still have two leading eigenvalues of different signs with the same magnitudes as before, as well as some very small eigenvalues. Raising $T'$ increases the small ones until they overtake the negative one. That is the moment when we consider that our dynamics is positive enough. In practice it is achieved by $T' \approx 0.3$ for norm of terms in the Hamiltonian $\|H_i\| \approx 1$. One can write down an explicitly positive evolution by keeping the above three eigenvaluese at this  $T'$ and dropping the rest, but the error one makes while keeping all the terms is small enough. And using the original form with $A$ and $A^f$ is so much simpler.

\paragraph{Numerical investigation} The matrix $\rho'G$ has small negative eigenvalues $\lambda$. One needs to be careful as the basis is not orthonormal, large norms may appear.
%we know that simple one-operator noise that corresponds to keeping just second eigenvalue models a contact with some infinite temperature bath
The hamiltonian is:
\begin{equation}
   H= 0.5 \sigma_z^1 -0.7  \sigma_z^2 + 0.3\sigma_z^1 \sigma_z^2 + \sigma_x^1 +\sigma_x^2 
\end{equation}
The $A=0.5\sigma_z^1$. Our investigation of positivity will apply to each portion of the noise separately, for small enough $dt$.  $A^f$ is taken to be $\int_{-\infty}^\infty$.

With that, we evaluate the eigenvalues of $\rho'$ for $dt =0.01$ for different $T'$.
\begin{figure}[h!]
\centering
\includegraphics[scale=0.6]{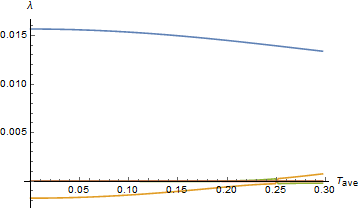}
\caption{Second eigenvalue in gold and third in green, as $T'$ is changed from $0$ to $0.3$}
\label{fig:full}
\end{figure}
%Tis a crossing lo.
\begin{figure}[h!]
\centering
\includegraphics[scale=0.6]{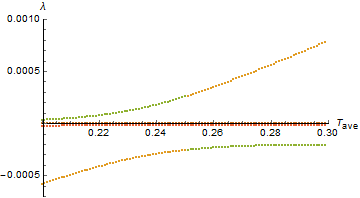}
\caption{Zoom in into the crossing of 3rd (lowest), 4th (highest) and 5th eigenvalues. $T'$ ranges from $0.2$ to $0.3$. The color scheme is due to reordering in absolute value.}
\label{fig:cross}
\end{figure}
It is an avoided crossing, so one can convince oneself that the level that used to be negative passes all the zeros and becomes positive.

\acknowledgments The author is thankful to Alexei Kitaev, John Preskill and Leonid Pryadko for useful comments and discussion.  E.M. carried out some of this work while supported by the Simons Foundation.

\appendix

\section{Exact time-averaging}
\label{app:ta}

Let's first formulate the steps of time averaging of a linear differential equation that can lead to sum of $T'$, $0$ and $-T'$ terms that appears in the Local Master Equation. We start with the equation
\begin{equation}
    \rho'(t) = M(t) \rho(t)
\end{equation}
where $M(t)$ contains rapidly changing terms. If we naively replace it by our desired equation
\begin{equation}
    \rho_1'(t) = \frac{1}{3}(M(t-T') + M(t) + M(t+T'))\rho_1(t)
\end{equation}
Then the difference $\|\rho -\rho_1\|$ becomes $\sim \frac{\|M'\|}{\|M\|}T'$ over 1 half-life. As $M$ contains both rapid and slow terms, we don't expect $\frac{\|M'\|}{\|M\|}$ to be small. It will be of order $\|H\|$ - the typical timescale of the hamiltonian. So
\begin{equation}
    \|\rho -\rho_1\|\approx \|H\| T' 
\end{equation}
In our case $\|H\| =1$ and $T'=0.26$ for 1 spin, which is too much of an error. A thermal state differs from maximally mixed state by roughly that much for $\beta =1$. We want to do better than that. We note that $\rho(t)$ will also contain rapid terms. We don't intend to keep track of oscillations faster than frequency $1/T'$, so we may as well consider evolution of the averaged density matrix
\begin{equation}
    \rho_{ave}(t) = \frac{1}{3}(\rho(t-T') + \rho(t) + \rho(t+T'))
\end{equation}
Note that the difference $\|\rho - \rho_{ave}\|$ is small at all times:
\begin{align}
    \rho_{ave}(t) = \frac{1}{3}(\rho(t-T') + \rho(t) + \rho(t+T')) =\\= \rho(t) +\frac{1}{3}\sum_{\Delta t} \delta\rho(\Delta t) \quad \quad \|\delta \rho\| \sim \|A\|^2 T'
\end{align}
Now we will derive the equation on $\rho_{ave}$, and see that it differs from a desired Markovian equation by a small amount. The difference between true $\rho_{ave}$ and the solution of our equation will be $\|A\|^2\beta T' $, which is tolerable for small $\|A\|$.
\begin{align}
    \rho_{ave}(t)' =\frac{1}{3}\sum_{\Delta t} M(t+ \Delta t) \rho(t+\Delta t) = \\=\frac{1}{3}\sum_{\Delta t} M(t+ \Delta t) \rho(t)+\\+\frac{1}{3}\sum_{\Delta t} M(t+ \Delta t)\delta\rho(\Delta t)=\\=\frac{1}{3}\sum_{\Delta t} M(t+ \Delta t) (\rho_{ave}(t) -\\-\frac{1}{3}\sum_{\Delta t'} \delta\rho(\Delta t')) +\frac{1}{3}\sum_{\Delta t} M(t+ \Delta t)\delta\rho(\Delta t)=\\ = 
    \frac{1}{3}\sum_{\Delta t} M(t+ \Delta t) \rho_{ave}(t) -\\-\frac{1}{3}\sum_{\Delta t} M(t+ \Delta t)\frac{1}{3}\sum_{\Delta t'} \delta\rho(\Delta t')) +\\+\frac{1}{3}\sum_{\Delta t} M(t+ \Delta t)\delta\rho(\Delta t) 
\end{align}
Here $\Delta t, \Delta t'$ range in $T',0,-T'$. The first term in the last line is the desired equation, the other are of order  $\|M\|\|\delta\rho\| \sim \|M\|^2 T'$. For a total evolution time $T$ such that $\|M\| T =1$ we get error $\|M\|T' = \|A^2\| \textrm{max}(t_b,\beta) T' $.

So if we find a solution of the above equation $\rho_{ave}(t)$ we are guaranteed that the solution of the original equation $\rho(t)$ is within $\delta \rho$ of it. In this way, we have proven that there is a tighter bound on $\|\rho - \rho_1 \|$ that we naively expected. Same applies for Heisenberg $\rho$ and $\rho_{ave}$, because the error term just gets unitary rotated, but does not increase in the norm.

Of course, the above trick can also be applied to the integral time averaging $\int_{-T'}^{T'} $.

Now let's try to extend this method of deriving equation on $\rho_{ave}$ to the integral equation. There's a freedom how to do that. We want all the $\rho$'s involved to be within $[-\infty, t]$, which can be done by $T'$ shifts in $\rho(t)$. An equation will then contain terms like (in the Heisenberg picture):
\begin{equation}
    \int_0^\infty A(T'-\tau)C(\tau) \rho_{ave}(t-\tau) d \tau A(T')
\end{equation}
which contain shifted $\rho_{ave}$ to prevent looking in the future, as well as terms
\begin{equation}
    \int_0^\infty A(-T'-\tau)C(\tau) \rho_{ave}(t-T'-\tau) d \tau A(-T')
\end{equation}
we can remove $T'$ from $\rho_{ave}$ for the sake of symmetry of the expression - that does not lead to a significant error. The counterterms also get their time-averaged version looking just like the Local Master Equation.

\begin{align}
  \frac{d\rho_{s}}{dt} = - i[H_s,\rho_s(t)] + \frac{1}{3} (
 [\int_0^\infty A(-\tau)C(\tau) \rho_s(t-\tau) d \tau ,A] + \\+ [A,\int_0^\infty A(-\tau)C^*(\tau) \rho_s(t-\tau) d \tau]+  \\
  +  [    \int_0^\infty A(T'-\tau)C(\tau) \rho_s(t-\tau) d \tau,A(T')] + \\+ [A(T'),    \int_0^\infty A(T'-\tau)C^*(\tau) \rho_s(t-\tau) d \tau] \\
  +  [\int_0^\infty A(-T'-\tau)C(\tau) \rho_s(t-T'-\tau) d \tau,A(-T')] + \\+ [A(-T'),\int_0^\infty A(-T'-\tau)C(\tau)^* \rho_s(t-T'-\tau) d \tau] )-\\- iD(0)( [A\rho_{s}(t),A] -  [A,\rho_{s}(t)A]+ \label{FinalForm2}  \\
  +  [A(T')\rho_{s}(t),A(T')] -  [A(T'),\rho_{s}(t)A(T')] \\
  +  [A(-T')\rho_{s}(t),A(-T')] -  [A(-T'),\rho_{s}(t)A(-T')] ))
\end{align}

\section{Bath correlation}
\label{app:b}
The master equation used the following correlation function of the environment: 
\begin{equation}
    \textrm{Tr} \rho_b B(t)B(\tau) = C(t-\tau)
\end{equation}
and its complex conjugate $C^* = \textrm{Tr} \rho_b B(\tau)B(t)$ if $B$ is a Hermitean operator. 

If it happens that $C(t)$ is purely real for all $t$, then the bath is infinite temperature. The imaginary part is necessary for finite temperature steady state. %In fact, if $C = C' + i C''$, then if we demand all two level systems to relax to thermal states, the condition we get is:
%\begin{equation}
%    \textrm{tanh} \beta E = \frac{\int_0^{\infty} d\tau C''(\tau) \textrm{sin} Et}{\int_0^{\infty} d\tau C'(\tau) \textrm{cos} Et} \label{CondOnC}
%\end{equation}
%Below we will confirm that any correlation function that stems from Gibbs ensemble does indeed satisfy that ( unless there are strategic zeroes in the matrix elements of $B$, then there may be a division by zero on the way to the above equation).

By a straightforward use of eigenbasis of the bath, we get
\begin{align}
C(t) = \textrm{Tr} \rho_b B(t)B = \sum_{nm} \rho_n e^{iE_n t}B_{nm}e^{-iE_m t}B_{mn} = \\
=\sum_{nm} \rho_n |B_{nm}|^2 e^{iE_{nm}t} = \int_{-\infty}^{\infty} S(\omega) e^{i\omega t}
\end{align}
Where
\begin{equation}
    S(\omega) = \sum_{nm} \rho_n |B_{nm}|^2 \delta(\omega - E_{nm})
\end{equation}
Note that $S(\omega)\geq 0$. Also, consider two delta functions, corresponding to $E_{mn}$ and $E_{nm} = - E_{mn}$, with amplitudes $\sim\rho_n \sim e^{-\beta E_n}$ and $\sim \rho_m\sim e^{-\beta E_m}$ correspondingly. The magnitude of $\delta$-function in $S(\omega)$ at $\omega = E_{mn}$ is $e^{-\beta E_{mn}} = e^{-\beta \omega}$ times the magnitude at $-\omega = E_{nm}$. This holds for all contributing $
\delta$ functions, thus it holds for appropriate smooth version of $S(\omega)$
 in the thermodynamic limit: 
 \begin{equation}
     S(\omega) = e^{-\beta \omega}S(-\omega) \label{CondOnS}
 \end{equation}
 
%If one follows through with definitions of $C(t)$, one can find that the above is equivalent to demanding the condition in equation (\ref{CondOnC}).

So we can take any $S(\omega)$ that is positive and satisfies Eq. (\ref{CondOnS}). From the derivation of the Master equation, there's an extra requirement that $C(t)$ decays with $t$ sufficiently fast. Besides that, $S$ can be any real positive function.

%Let's derive the equivalence of conditions on $S$ and $C$. Since $S(\omega)$ is real, not complex, the corresponding Fourier transform $C(t)$ possesses the property $C(t)^* = C(-t)$ if extended to negatives. Using that, we get 
%\begin{equation}
%   S(\omega) = \int_0^{\infty}C(t) e^{-i\omega t} + \int_0^{\infty}C(t)^* e^{i\omega t}
%\end{equation}
%After some cancellations:
%\begin{equation}
%   S(\omega) = 2\int_0^{\infty}C'(t) \textrm{cos}\omega t + 2\int_0^{\infty}C''(t) \textrm{sin}\omega t
%\end{equation}
%Now plugging that into $ S(\omega) = e^{\beta \omega}S(-\omega)$ leads to the condition with hyperbolic tangent, while $S(\omega)>0$ combined with previous one gives the Positivity condition below:
%\begin{equation}
%   \forall \omega \quad  S(\omega) >0 \Leftrightarrow \forall \omega \quad (1 + \textrm{tanh} \beta \omega) \int_0^{\infty}C'(t) \textrm{cos} \omega t >0
%\end{equation}

%Note that it is easier to guess $S(\omega)$ with the decay of the Fourier transform, than guessing $C(t)$ that satisfy obscure conditions above. I just wanted to stress out that they are not too restrictive. However, naive complex choices like $e^{-\lambda_1 t} + i e^{-\lambda_2 t}$ will not work because they do not correspond to any one temperature.

There's one particularly simple choice $S(\omega) = e^{-\omega^2/\sigma - \beta \omega/2}$.  Let's calculate the corresponding decay time, as this is the one we will use in the numerics.
\begin{align}
    C(t) = \int_{-\infty}^{\infty}S(\omega) e^{i\omega t}d\omega= \sqrt{\sigma \pi} e^{-\frac{\sigma}{4}(t-i\beta/2)^2 } =\\= \sqrt{\sigma \pi} e^{\sigma \beta^2/16 -\sigma t^2/4 + i\sigma\beta t/4} 
\end{align}
We see that the decay time is:
\begin{equation}
    t_{b} =\sqrt{\frac{2\textrm{ln}\pi\sigma}{\sigma} + \beta^2}
\end{equation}
So we need both high $T$
 and broad (fast) bath for the derivation of Markovian master equation to be valid.

\onecolumngrid
\section{True Fixed Point}
\label{true}

\paragraph{Derivation of fixed point equations}

\begin{align}
C(t) = \textrm{tr}_B \rho_{G,b}B(t) B= \sum_{nm}\rho_n^{G,b}|B_{nm}|^2 e^{itE_{nm}} \\
C(\tau-\tau'>0) =\textrm{tr}_B \rho_{G,b}B(\tau) B(\tau')= \sum_{nm}\rho_n^{G,b}|B_{nm}|^2 e^{(\tau-\tau')E_{nm}} 
\end{align}
After splitting $\sum_{nm} = \sum_E \sum_{nm: E_nm=E}$ and denoting 
\begin{equation}
     X(E) = \sum_{nm: E_{nm}=E}\rho_n^{G,b}|B_{nm}|^2
\end{equation}
the correlation functions can be related by analytic continuation in the following way
\begin{align}
     \sum_E X(E) \delta(E-\omega) = S(\omega)\\
      \sum_E X(E)e^{iEt} = \int_{-\infty}^\infty e^{i\omega t}\sum_E X(E) \delta(E-\omega) d\omega = \int_{-\infty}^\infty e^{i\omega t} S(\omega) d\omega = C(t)\\
       \sum_E X(E)e^{E\tau} = \int_{-\infty}^\infty e^{\omega \tau}\sum_E X(E) \delta(E-\omega) d\omega = \int_{-\infty}^\infty e^{\omega \tau} S(\omega) d\omega=C(\tau)
\end{align}
We also note that thanks to the Gibbs factor the $X(E)$ possesses the property
\begin{equation}
     X(E) = e^{-\beta E_{nm}}X(-E) \quad \Rightarrow \quad S(\omega) =e^{-\beta \omega} S(-\omega)
\end{equation}
Now let's express $A_f$:
\begin{equation}
     A_f = \int_{-\infty}^0 A(t) C(-t) dt= \int_{-\infty}^\infty A(t) e^{-i\omega t}S(\omega) dt d\omega
\end{equation}
Now 
\begin{equation}
     A(t) = \sum_{nm} A_{nm}e^{iE_{nm}t +\epsilon t }
\end{equation}
where small $\epsilon$ was added to keep track of the pole shifts below.
\begin{equation}
     A_f = \sum_{nm}A_{nm}\int_{-\infty}^\infty \frac{S(\omega)d\omega}{-i\omega +iE_{nm} +\epsilon}  =  i\sum_{nm}A_{nm}\int_{-\infty}^\infty \frac{S(\omega)d\omega}{\omega -E_{nm} +i\epsilon}
\end{equation}
Now the integral splits into the principal value and the pole:
\begin{equation}
   \int_{-\infty}^\infty \frac{S(\omega)d\omega}{\omega -E_{nm} +i\epsilon}= \textrm{p.v.} \int_{-\infty}^\infty \frac{S(\omega)d\omega}{\omega -E_{nm}} - \pi i S(E_{nm})
\end{equation}
so
\begin{equation}
     A_f =   \sum_{nm}A_{nm}(i\textrm{p.v.} \int_{-\infty}^\infty \frac{S(\omega)d\omega}{\omega -E_{nm}} + \pi  S(E_{nm}))
\end{equation}

Now let's look at the Gibbs preservation. The environment-induced part of the equation is as follows:
\begin{equation}
     \rho_{G,s}'= A^{f}\rho_{G,s}A-AA^{f}\rho_{G,s} +  A\rho_{G,s}A^{{f}\dag}-\rho_{G,s}A^{{f}\dag}A
\end{equation}
We need to see what it does to Gibbs state. First investigate the component of $A_f$ that is proportional to $S$. Note that the Hermitean conjugation flips the sign of the energy:
\begin{equation}
     A_f =  \sum_{nm}A_{nm}(\dots + \pi  S(E_{nm})), \quad A_f^\dag =  \sum_{nm}A_{nm}(\dots + \pi  S(E_{mn}))
\end{equation}
We arrive at %{\color{red}indices but no bra-ket}:
\begin{equation}
     \rho_{G,s, nm}'= \pi\sum_{nkm}A_{nk} A_{km} ( S(E_{nk})\rho^{G,s}_k -S(E_{km})\rho^{G,s}_m +  S(E_{mk})\rho^{G,s}_k -S(E_{kn})\rho^{G,s}_n) +\dots
\end{equation}
we group the first and the last term, and the middle two:
\begin{equation}
     \rho_{G,s,nm}'= \pi\sum_{nkm}A_{nk} A_{km} ( (S(E_{nk})\rho^{G,s}_k -S(E_{kn})\rho^{G,s}_n) -(S(E_{km})\rho^{G,s}_m -  S(E_{mk})\rho^{G,s}_k )) +\dots
\end{equation}
One may observe that they cancel each other as grouped:
\begin{align}
     S(E_{nk})\rho^{G,s}_k -S(E_{kn})\rho^{G,s}_n=0 \\
    S(E_{km})\rho^{G,s}_m -  S(E_{mk})\rho^{G,s}_k=0
\end{align}
because
\begin{equation}
         S(E_{nk})\rho^{G,s}_k -S(E_{kn})\rho^{G,s}_n= \rho^{G,s}_k(S(E_{nk}) -S(E_{kn})e^{\beta E_{kn}})=\rho^{G,s}_k(S(\omega) -S(-\omega)e^{-\beta \omega})=0
\end{equation}
 where we have used the special property of spectral density. So the equation with just $S$ terms would preserve the isolated system Gibbs state. Now let's derive the remaining terms containing principal value integrals. For brevity, we denote
\begin{equation}
     D(E_{nm}) = \textrm{p.v.} \int_{-\infty}^\infty \frac{S(\omega)d\omega}{\omega -E_{nm}}
\end{equation}
so the part of $A_f$ we care about is (let's call it $A_{p.v.}$)
\begin{equation}
     A_{p.v.} =   \sum_{nm}A_{nm}iD(E_{nm})
\end{equation}
Now the full expression for Hermitean conjugation can be written concisely:
\begin{equation}
     A_f =  \sum_{nm}A_{nm}(iD(E_{nm})+ \pi  S(E_{nm})), \quad A_f^\dag =  \sum_{nm}A_{nm}(-iD(E_{mn}) + \pi  S(E_{mn}))
\end{equation}
We calculate the remaining part of the evolution of the Gibbs state:
\begin{equation}
     \rho_{G,s,nm}'= \dots +i\sum_{nkm}A_{nk} A_{km} ( D(E_{nk})\rho^{G,s}_k -D(E_{km})\rho^{G,s}_m -  D(E_{mk})\rho^{G,s}_k +D(E_{kn})\rho^{G,s}_n)
\end{equation}
this part does not, in general, vanish. We will see how it contributes to the true fixed point below.

\paragraph{True Gibbs preservation} 
%One just needed to be more careful with finite elements integration formula used in the Integral equation.
The Integral equation decays to the same fixed point as the Markovian equation, as it should because fixed point condition involves time-independent $\rho$:
\begin{equation}
      \frac{d\rho_{s}}{dt} =0 = - i[H_s,\rho_{G,s}+\delta\rho] +   [A^{f}(\rho_{G,s}+\delta\rho),A] +  [A,(\rho_{G,s}+\delta\rho)A^{{f}\dag}]
\end{equation}
In perturbation theory, that fixed point is shifted by $\sim A^2$ terms away from the isolated system Gibbs distribution. We do not know if there's a general bound on the size of those terms, and thus on the validity of the perturbative approach. But we will see that it is valid at least in a sense that the matrix element deviations $\delta \rho$ are small under not very restrictive conditions.
%for the straightforward calculation of tr$_B \rho_{total,G}$.For the fixed point achieved in this equation, the calculation is doable. It involves counterterms from Lindblad on diagonal, as
For a perturbative calculation, one notes that $\rho_{G,s}$ commutes with the Hamiltonian, and that $\delta\rho$ can be dropped in the decay term in the first order calculation: 
\begin{equation}
    i[H_s,\delta\rho] =   [A^{f}\rho_{G,s},A] +  [A,\rho_{G,s}A^{{f}\dag}]
\end{equation}
We have already done the calculation for the terms on the right:
\begin{equation}
  i[H_s,\delta\rho]_{nm}=   i\sum_{nkm}A_{nk} A_{km} ( D(E_{nk})\rho^{G,s}_k -D(E_{km})\rho^{G,s}_m -  D(E_{mk})\rho^{G,s}_k +D(E_{kn})\rho^{G,s}_n)
\end{equation}

In the first order, the fixed point has the same diagonal (one can easily see that by checking that $n=m$ terms cancel from the above sum). The offdiagonal contributions will be (new signs!):
\begin{equation}
    \delta \rho_{nm} = \sum_k A_{nk}A_{km}\frac{ e^{-\beta E_k}  (D(E_{nk}) - D(E_{mk})) -  e^{-\beta E_m}D(E_{km}) +e^{-\beta E_n}D(E_{kn}) }{ZE_{nm}}
\end{equation}

%\begin{equation}
%    \delta \rho_{nm} = \sum_k A_{nk}A_{km}\frac{ e^{-\beta E_k}  (D(E_{nk}) + D(E_{km})) -  e^{-\beta E_m}D(E_{km}) -e^{-\beta E_n}D(E_{nk}) }{ZE_{nm}}
%\end{equation}
As usual perturbation theory calculation, it doesn't behave well at $E_{nm}\to 0$, and degenerate perturbation theory should be used. We note that neither of cases ($E_{nm}=0, E_{nm}>\Delta E$) provides a good bound on $\delta \rho$ in general, but under certain assumptions it is small.

%More useful Mathematica links on plotting:

%https://reference.wolfram.com/language/tutorial/RedrawingAndCombiningPlots.html

%https://reference.wolfram.com/language/howto/CombineTwoOrMoreGraphics.html

%https://reference.wolfram.com/language/tutorial/TypingSubscripts.html

More specifically, for a system of size $L$ with minimal spacing $\Delta E$ we can bound the $\delta \rho$ at least as:
\begin{equation}
    \|\delta \rho\| \leq \frac{\textrm{exp}(L) A^2}{\Delta E}
\end{equation}
using the above formula.

Another thing one may wish to prove is that $\delta \rho$, if not small, is at least local in some sense. And indeed, if the system decays at a good rate, the equation itself is the proof that it is local. For systems that do not relax well due to big memory times the naive evolution with our equation does not prepare a thermal state, neither we expect it to be related approximately by a local superoperator to the original one.% Indeed, if the noise induces a constant Lamb shift in a ferromagnet ground states, the Gibbs state is changed everywhere (for temperatures in the ferromagnetic phase). So that is an example of small in norm (every eigenstate probability gets shifted by small amount), but nonlocal change in fixed point.

Now let's compare our fixed point to the true reduced density matrix
\begin{equation}
    \rho_{true} =\frac{1}{Z} \textrm{tr}_B E^{-\beta(H_s +V +H_b)} = \rho_{G,s} + \textrm{tr}_B\rho_{G,s}\otimes \rho_{G,b} \int \int_{\tau>\tau'} d\tau d\tau' V(\tau) V(\tau')  + O(V^4)
\end{equation}
the expansion is not neccessarily valid, because norms $\|V(\tau)\|$ for $\tau\in[0,\beta]$ do not have to be of the same order as $\|V\|$. There may be huge $e^{\beta \Delta E}$ factors in them. Maybe if one carefully investigates this expression, one will find suppression anyways, because of the $\rho$ factors. There don't seem to be an easy counterexample. %In fact, a no imaginary time LRB counterexample will do.

Surprisingly, the above formula is exactly the fixed point of our equation! So to this order $\rho_{true} = \rho_{G,s} + \delta \rho$ where $\delta \rho$ is the same as one found by  perturbation theory of the fixed point condition. It also helps to figure out the fate of apparent zero in the denominator in that equivalence. In fact, one of the integrals turns out to give that:
\begin{equation}
    \int_{\alpha/2}^{\beta -\alpha/2} d(\frac{\tau + \tau'}{2}) e^{E_{nm} (\frac{\tau+\tau'}{2})}  = \frac{e^{ E_{nm}\alpha/2}-e^{ E_{nm}(\beta-\alpha/2)}}{E_{nm}}
\end{equation}
where $\alpha = \tau - \tau'$. Then two terms in the numerator lead to different summands in expression for $\delta \rho$. In other words, appropriate grouping of the term should remove the apparent pole.

So we are convinced that our equation (both Markovian and Integral) has true reduced density matrix tr$_B e^{-\beta(H_s + V + H_b)}/Z$ as a fixed point. There's nothing more physical than that. If one demands the Gibbs state of $H_s$, we can't really guarantee to be close to it in the norm. Every matrix element will be close to the one of isolated system as $A^2$ (if we forget about the pole). Neither we know how to introduce counterterms that will restore it as an exact fixed point.

We will present a suggestion that works for special systems - many-body localized ones. There one indeed can hope for corrections to be small and local. A little bit more generally, one may be able to bound the correction for any locally generated fixed point (the one that can be disentangled by local operations across any cut).

\paragraph{The derivation}
%The signs are taken from 
%http://www.physics.ucla.edu/~nayak/many_body.pdf
\begin{equation}
    \delta \rho_{nm}^{true}=\textrm{tr}_B \rho_{G,s}\otimes \rho_{G,b}\int \int_{\tau>\tau'} d\tau d\tau' V(\tau) V(\tau')   =  \rho_{G,s}\int \int_{\tau>\tau'} d\tau d\tau' A(\tau) A(\tau')C(\tau -\tau')
\end{equation}
where $C(\tau) = \textrm{tr}_B \rho_{G,b} B(\tau) B(\tau')$ is an analytic continuation of $C(t)$. Explicitly we had:
\begin{align}
C(t) = \sum_{nm}\rho_n^{G,b}|B_{nm}|^2 e^{itE_{nm}} = \int_{-\infty}^\infty S(\omega) e^{i\omega t}\\
C(\tau) = \sum_{nm}\rho_n^{G,b}|B_{nm}|^2 e^{(\tau-\tau')E_{nm}} = \int_{-\infty}^\infty S(\omega) e^{\omega(\tau-\tau')}
\end{align}
Let's do the same eigenstate decomposition on the system now:
\begin{equation}
     \delta \rho_{nm}^{true}=\rho_n^{G,s}\sum_k A_{nk}A_{km}\int \int_{\tau>\tau'}^\beta d\tau d\tau' e^{\tau E_{nk}} e^{\tau'E_{km}}\int_{-\infty}^\infty S(\omega)e^{(\tau -\tau')\omega}d\omega
\end{equation}
Recall that in our target formula
\begin{equation}
     D(E) = -\textrm{p.v.} \int_{-\infty}^\infty d\omega \frac{S(\omega) d\omega}{E-\omega}
\end{equation}
The non-principal value part corresponding to the pole is also there, but it cancels due to thermal law. We will try to collect the above expressions in the result of integration over $\tau, \tau'$. First we shift to new variables $\alpha = \tau- \tau'$ and $\gamma = (\tau+\tau')/2$. The integrals become:
\begin{align}
     \int_0^\beta d\alpha \int_{\beta-\alpha/2}^{ \alpha/2}e^{\gamma E_{nm}}d\gamma e^{\frac{\alpha}{2}(E_{nk}+E_{mk} +2\omega)} =\\= -\int_0^\beta d\alpha \frac{e^{\alpha E_{nm}/2}-e^{\beta E_{nm}}e^{-\alpha E_{nm}/2}}{E_{nm}} e^{\frac{\alpha}{2}(E_{nk}+E_{mk} +2\omega)} =\\= -
     \int_0^\beta d\alpha \frac{e^{\alpha (E_{nk}+\omega)}-e^{\beta E_{nm}}e^{-\alpha (E_{mk}+\omega)}}{E_{nm}} =\\=\frac{1}{E_{nm}}\left(\frac{e^{\beta (E_{nk}+\omega)}-1}{E_{nk}+\omega} -e^{\beta E_{nm}}\frac{e^{\beta (E_{mk}+\omega)}-1}{E_{mk}+\omega} \right)     
\end{align}
It will be neccessary to use the property $S(\omega) = e^{\beta \omega}S(-\omega)$:  (recover factors!)
\begin{align}
     \delta \rho_{nm}^{true}=\\=\rho_n^{G,s}\sum_k \frac{A_{nk}A_{km}}{E_{nm}}\int_{-\infty}^\infty \left(\frac{e^{\beta (E_{nk}+\omega)}-1}{E_{nk}+\omega} -e^{\beta E_{nm}}\frac{e^{\beta (E_{mk}+\omega)}-1}{E_{mk}+\omega} \right)  S(\omega)d\omega =\\=\rho_n^{G,s}\sum_k \frac{A_{nk}A_{km}}{E_{nm}}\int_{-\infty}^\infty \left(\frac{e^{\beta (E_{nk}+\omega)}}{E_{nk}+\omega} -e^{\beta E_{nm}}\frac{e^{\beta (E_{mk}+\omega)}}{E_{mk}+\omega} \right)  S(\omega)d\omega  - \\-\rho_n^{G,s}\sum_k \frac{A_{nk}A_{km}}{E_{nm}}\int_{-\infty}^\infty \left(\frac{1}{E_{nk}+\omega} -e^{\beta E_{nm}}\frac{1}{E_{mk}+\omega} \right)  S(\omega)d\omega  =\\=\rho_n^{G,s}\sum_k \frac{A_{nk}A_{km}}{E_{nm}}\int_{-\infty}^\infty \left(\frac{e^{\beta E_{nk}}}{E_{nk}-\omega} -e^{\beta E_{nm}}\frac{e^{\beta E_{mk}}}{E_{mk}-\omega} \right)  S(\omega)d\omega  -\\- \rho_n^{G,s}\sum_k \frac{A_{nk}A_{km}}{E_{nm}} \left(-D(E_{kn}) +e^{\beta E_{nm}}D(E_{km}) \right) =\\= \rho_n^{G,s}\sum_k \frac{A_{nk}A_{km}}{E_{nm}} \left(e^{\beta E_{nk}}D(E_{nk}) -e^{\beta E_{nk}}D(E_{mk}) +D(E_{kn}) -e^{\beta E_{nm}}D(E_{km})\right) 
     %=\\=
     % \rho_n^{G,s}\sum_k \frac{A_{nk}A_{km}}{E_{nm}} \left(e^{\beta E_{nk}}D(E_{nk})+e^{\beta E_{nm}}D(E_{kn}) -e^{\beta E_{nk}}D(E_{mk})  -D(E_{km})\right)
\end{align}
%We didn't quite get the same answer... Need to debug.
Note that we implied the principal value because the terms with $iD(E) \to S(E)$ cancel due to thermal law. The signs are wrong for that though... Maybe there are strategic $\pm$ that fix it - I don't know how to keep track of pole bypass directions. The final answer for the correction is:
\begin{equation}
   \delta \rho_{nm}^{true}=  \sum_k \frac{A_{nk}A_{km}}{ZE_{nm}} \left(e^{-\beta E_{k}}D(E_{nk}) -e^{-\beta E_{k}}D(E_{mk}) +e^{-\beta E_{n}}D(E_{kn}) -e^{-\beta E_{m}}D(E_{km})\right) 
\end{equation}

Let's look at the other derivation -
%\begin{itemize}
%    \item $i$ cancels with $i[H,\rho]$
%    \item $e^{\beta E_k}$ are individual because that's how they appear thanks to the Gibbs distribution on the rhs
%    \item complex conjugation of $A_{p.v.}$ may or may not flip the argument Idk
%\end{itemize}

the answer we got there was
%\begin{equation}
 %      \frac{1}{Z}\sum_k \frac{A_{nk}A_{km}}{E_{nm}} \left(e^{-\beta E_{k}}D(E_{nk})+e^{-\beta E_{m}}D(E_{kn}) -e^{-\beta E_{k}}D(E_{mk})  -e^{-\beta E_{n}}D(E_{km})\right)
%\end{equation}
\begin{equation}
    \delta \rho_{nm} = \sum_k A_{nk}A_{km}\frac{ e^{-\beta E_k}  (D(E_{nk}) - D(E_{mk})) -  e^{-\beta E_m}D(E_{km}) +e^{-\beta E_n}D(E_{kn}) }{ZE_{nm}}
\end{equation}
Indeed perfect coincidence.

 In fact the above calculation of equivalence of fixed point to reduced Gibbs state only applies for $E_{nm}\ne 0$. For $E_{nm}= 0$ the reduced Gibbs state expression has to be used as if in the limit
 \begin{equation}
     \textrm{lim}_{E_{nm}\to0} \delta \rho
 \end{equation}
 That limit indeed exists, as can be seen from the expression directly:
 \begin{equation}
    \delta \rho_{nm} = \sum_k A_{nk}A_{km}\frac{ e^{-\beta E_k}  (D(E_{nk}) - D(E_{mk})) -  e^{-\beta E_m}D(E_{km}) +e^{-\beta E_n}D(E_{kn}) }{ZE_{nm}}
\end{equation}
The numerator is zero at $E_{nm}=0$ and its first derivative gives the following limit:
 \begin{equation}
    \delta \rho_{nm} = \sum_k A_{nk}A_{km}\frac{-\beta  e^{-\beta E_m}D'(E_{km})+ e^{-\beta E_k}   D'(E_{mk}) +\beta  e^{-\beta E_n}D(E_{kn}) }{Z}
\end{equation}
The last term also appears if one does the counterterm trick in the traditional Lindblad equation. We do not know of the interpretation of the first two. $D'$ doesn't satisfy the thermal law, so they don't cancel.

But our first order fixed point condition is trivially satisfied at $E_{nm}=0$, so any $O(A^2)$ change in diagonal elements of $\rho$ preserves the condition to that order. We assumed that there are no degeneracies in the Hamiltonian for simplicity, so we can call $E_{nm}=0$ the diagonal. In higher order we find a fixed point of the diagonal to $O(A^2)$ precision, and offdiagonal elements to $O(A^4)$ precision. The fixed point of traditional Lindblad equation can be changed by $O(A^2)$ amount by introducing appropriate counterterms, so it can be made to be exactly the diagonal of reduced Gibbs state. The offdiagonal fixed point is zero for Lindblad. The fixed point without counterterms is just $\rho_{G,s}$ of isolated system. 

In our case, there are offdiagonal terms in the fixed point, so they will shift the diagonal. Specifically, one can denote a superoperator:
\begin{equation}
    \mathcal{A} : \rho \to [A^{f}\rho,A] +  [A,\rho A^{{f}\dag}]
\end{equation}
The second order perturbation theory for fixed point is then:
\begin{equation}
    0=i[H,\delta \rho^{(2)}] + \mathcal{A}(\delta \rho^{(1)})
\end{equation}
where $\delta \rho^{(1)} = \delta \rho_E^{(1)} +\delta\rho_0^{(1)} $ where $\delta \rho_E^{(1)}$ is $E_{nm}\ne 0$ already found in the first, and $\delta\rho_0^{(1)}$ is the $E_{nm}= 0$ to be determined. By taking $E_{nm}= 0$ elements of the above equation, we kill the contributions of $\delta \rho^{(2)}$:
\begin{equation}
    0= \mathcal{A}_0( \delta \rho_E^{(1)} +\delta\rho_0^{(1)})
\end{equation}
or
\begin{equation}
  \mathcal{A}_0(\delta\rho_0^{(1)})=-\mathcal{A}_0( \delta \rho_E^{(1)} )
\end{equation}
So we just need to invert the operator $A_0$ that maps diagonal to diagonal. But that is exactly the relaxation part of the traditional Lindblad equation! It has zero eigenvalue corresponding to $\rho_{G,s}$, which is expected from perturbation theory (we won't find meaningful corrections to the amplitude of zeroth order state). But in the subspace of other eigenvectors we can invert this matrix:
\begin{equation}
  \delta\rho_0^{(1)}=-A_{00, \lambda\ne0}^{-1}\mathcal{A}_0( \delta \rho_E^{(1)} ) 
\end{equation}
or in components:
\begin{equation}
  \delta\rho^{(1)}_n =-\sum_{\lambda\ne 0} \lambda^{-1} v_n \sum_m v_m^*\mathcal{A}_{mm}( \delta \rho_E^{(1)} )
\end{equation}
where $v$ are eigenvectors of the Markov process. This form is a bit concerning because $\lambda^{-1}$ can be arbitrarily big, but we remember from our original estimates that if diagonal deviates more than $O(A^2)$ from the Gibbs state, the first order condition will not be able to balance them anymore, so fixed point can't be further than that. So we expect that by appropriate counterterms we can remove this correction and restore the true fixed point diagonal of reduced Gibbs state (or isolated Gibbs state, if one so desires).
 
\bibliographystyle{unsrt}
\bibliography{LastPush}

\end{document}